\documentclass[lettersize,journal]{IEEEtran}
\usepackage{amsmath,amsfonts}
\usepackage{algorithmic}
\usepackage{algorithm}
\usepackage{array}
\usepackage[caption=false,font=normalsize,labelfont=sf,textfont=sf]{subfig}
\usepackage{textcomp}
\usepackage{stfloats}
\usepackage{url}
\usepackage{verbatim}
\usepackage{graphicx}
\usepackage{cite}
\usepackage{amssymb}
\usepackage{hyperref}

\begin{document}

\title{VQ-CTAP: Cross-Modal Fine-Grained Sequence Representation Learning for Speech Processing}

\author{Chunyu Qiang, \emph{Student Member, IEEE,}
    Wang Geng,
    Yi Zhao,
    Ruibo Fu, \emph{Member, IEEE,}
    Tao Wang, \emph{Student Member, IEEE,}
    Cheng Gong, \emph{Student Member, IEEE,}
    Tianrui Wang, \emph{Student Member, IEEE,}
    Qiuyu Liu, \emph{Student Member, IEEE,}
    Jiangyan Yi, \emph{Member, IEEE,}
    Zhengqi Wen, \emph{Member, IEEE,}
    Chen Zhang,
    Hao Che,
    Longbiao Wang, \emph{Member, IEEE,}
    Jianwu Dang, \emph{Member, IEEE,}
    Jianhua Tao, \emph{Senior Member, IEEE,}
\thanks{This work was supported by the National Natural Science Foundation of China (No. U23B2053, 62176182, 62302333, 62322120, U21B2010, and U2436210)}

\thanks{Chunyu Qiang is with School of New Media and Communication, Tianjin University, Tianjin, China, and also with Kuaishou Technology Co., Ltd., Beijing, China. (e-mail: \url{qiangchunyu@tju.edu.cn})}
\thanks{Wang Geng, Yi Zhao and Chen Zhang are with Kuaishou Technology Co., Ltd., Beijing, China.}
\thanks{Ruibo Fu and Tao Wang are with Institute of Automation Chinese Academy of Sciences, Beijing, China. }
\thanks{Cheng Gong, Tianrui Wang, Qiuyu Liu and Longbiao Wang are with Tianjin Key Laboratory of Cognitive Computing and Application, College of Intelligence and Computing, Tianjin University, Tianjin, China.  }
\thanks{Jiangyan Yi and Jianhua Tao are with Department of Automation, Tsinghua University, Beijing, China.}
\thanks{Zhengqi Wen is with Beijing National Research Center for Information Science and Technology, Tsinghua University, Beijing, China.}
\thanks{Hao Che is with Migu Culture Technology Co., Ltd., Beijing, China.  }
\thanks{Jianwu Dang is with Shenzhen Institute of Advanced Technology, Chinese Academy of Science, Guangdong, China.}
\thanks{Longbiao Wang is the corresponding author. (e-mail: \url{longbiao_wang@tju.edu.cn})}
}
\markboth{Journal of \LaTeX\ Class Files}%
{Shell \MakeLowercase{\textit{et al.}}: VQ-CTAP: Cross-Modal Fine-Grained Sequence Representation Learning for Speech Processing}


\maketitle

\begin{abstract}
Deep learning has brought significant improvements to the field of cross-modal representation learning. For tasks such as text-to-speech (TTS), voice conversion (VC), and automatic speech recognition (ASR), a cross-modal fine-grained (frame-level) sequence representation is desired, emphasizing the semantic content of the text modality while de-emphasizing the paralinguistic information of the speech modality. We propose a method called ``Vector Quantized Contrastive Token-Acoustic Pre-training (VQ-CTAP)", which uses the cross-modal aligned sequence transcoder to bring text and speech into a joint multimodal space, learning how to connect text and speech at the frame level. The proposed VQ-CTAP is a paradigm for cross-modal sequence representation learning, offering a promising solution for fine-grained generation and recognition tasks in speech processing. The VQ-CTAP can be directly applied to VC and ASR tasks without fine-tuning or additional structures. We propose a sequence-aware semantic connector, which connects multiple frozen pre-trained modules for the TTS task, exhibiting a plug-and-play capability. We design a stepping optimization strategy to ensure effective model convergence by gradually injecting and adjusting the influence of various loss components. Furthermore, we propose a semantic-transfer-wise paralinguistic consistency loss to enhance representational capabilities, allowing the model to better generalize to unseen data and capture the nuances of paralinguistic information. In addition, VQ-CTAP achieves high-compression speech coding at a rate of 25Hz from 24kHz input waveforms, which is a 960-fold reduction in the sampling rate. The experimental results demonstrate that while VQ-CTAP outperforms baseline methods in TTS and VC tasks, its performance on ASR tasks is suboptimal. The audio demo is available at \url{https://qiangchunyu.github.io/VQCTAP/}.
\end{abstract}


\begin{figure}[ht]
  \centering
  \includegraphics[width=\linewidth]{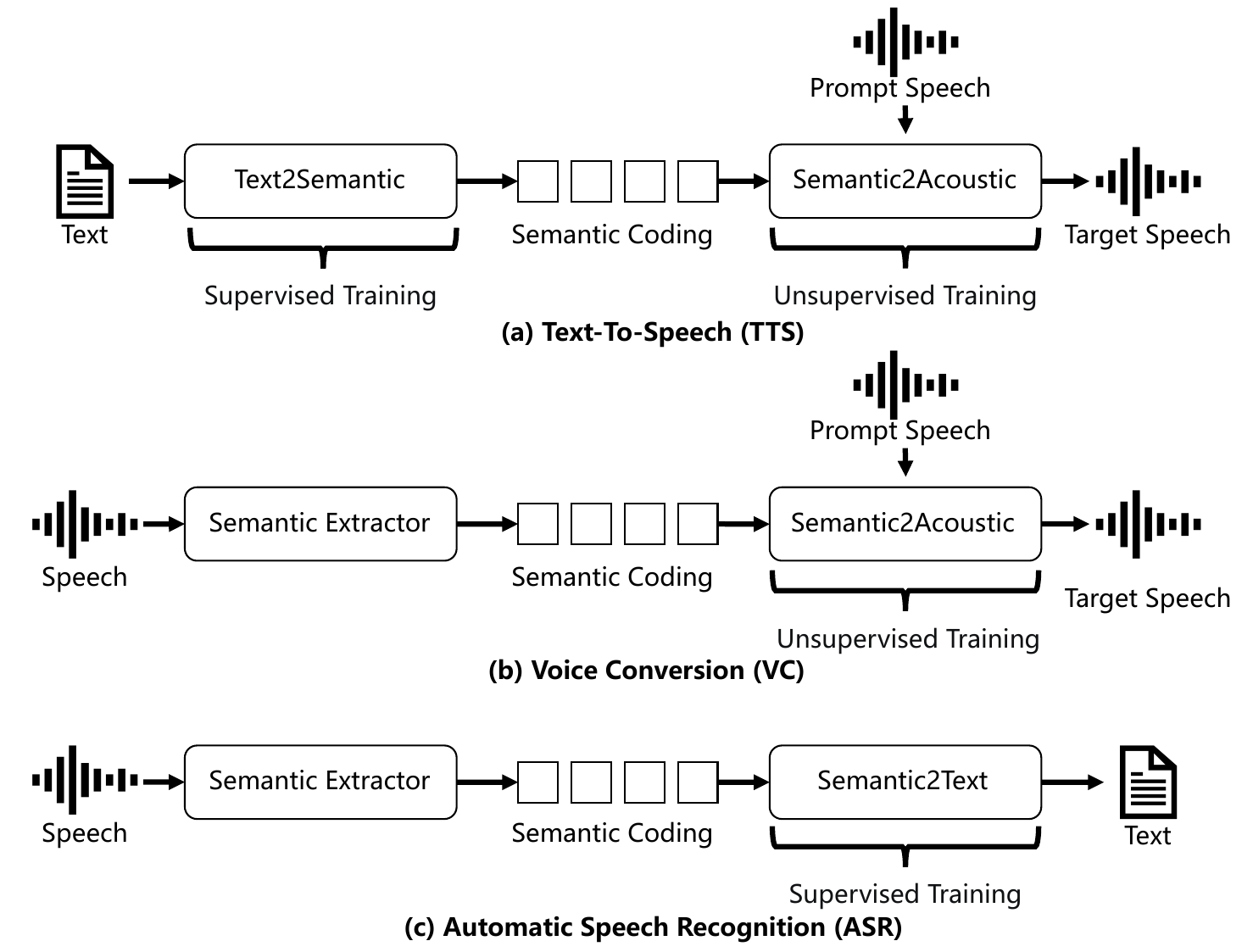}
  \caption{TTS, VC and ASR systems based on semantic coding}
  \label{fig:tts-vc-asr}
\end{figure}

\section{Introduction}
\IEEEPARstart{D}{eep} learning has brought significant improvements to the field of cross-modal representation, leading to the emergence of numerous methods for representing connections between text, images, and speech\cite{radford2021learning, yuan2021florence, jia2021scaling, wu2022wav2clip, guzhov2022audioclip, elizalde2023clap, girdhar2023imagebind, zhu2023languagebind}.

Contrastive learning has proven to be an effective method for cross-modal modeling. This approach has been widely applied in text-image cross-modal representation, with notable examples including OpenAI's CLIP\cite{radford2021learning}, Florence\cite{yuan2021florence}, and ALIGN\cite{jia2021scaling}. In the field of text-audio cross-modal representation, CLIP-based models have also been developed, including Wav2CLIP\cite{wu2022wav2clip}, AudioCLIP\cite{guzhov2022audioclip}, and CLAP\cite{elizalde2023clap}.  In the field of multi-modal representation, pre-trained representation models have emerged that use a single modality as a link to connect other modalities. For example, ImageBind\cite{girdhar2023imagebind} achieves joint modeling of multiple modalities based on the image modality, while LanguageBind\cite{zhu2023languagebind} achieves joint modeling of multiple modalities based on the language modality. These models focus on extracting global descriptive information; for instance, CLAP aims to enhance performance in downstream audio classification tasks. However, their emphasis on global information limits their applicability to tasks requiring fine-grained sequence generation and recognition.

As shown in Figure \ref{fig:tts-vc-asr}, the speech processing tasks of TTS and ASR are typical cross-modal applications. For tasks such as TTS, VC and ASR, a fine-grained (frame-level) sequence cross-modal representation(semantic coding) between text-speech modalities is expected. Semantic should provide a representation of speech where the linguistic content— from phonetics to semantics—is salient, while paralinguistic information such as speaker identity and acoustic details are removed\cite{kharitonov2023speak}. The intermediate representation extracted from speech should serve as a ``bridge" between text and acoustic information. It should emphasize linguistic content while de-emphasizing paralinguistic information such as speaker identity and acoustic details. Self-supervised representation learning methods, such as Wav2Vec2.0\cite{baevski2020wav2vec}, Wav2Vec-C\cite{sadhu2021wav2vec}, VQ-Wav2Vec\cite{baevski2019vq}, HuBERT\cite{Hsu2021HuBERTSS}, and W2V-BERT\cite{chung2021w2v}, have become mainstream in the speech representation field. These methods offer the prospect of a universal model that can benefit a wide range of tasks and domains. However, when applied to VC and TTS tasks, such as SPEAR-TTS\cite{kharitonov2023speak}, these methods encounter issues of redundancy and dimensionality explosion. This is because the lack of explicit text supervision in these methods results in learned representations that are not fully decoupled semantic information. Supervised frame-level representation learning methods aim to explicitly introduce text information as supervision. For example, Phonetic Posteriorgrams (PPGs)\cite{sun2016phonetic} are calculated based on ASR acoustic models. Although PPGs have been widely used in various speech processing tasks, they are essentially text information, and the frame-level features are independent of each other, lacking the integration of contextual semantic information. Moreover, these methods rely on expensive annotated text-speech pairs. In this case, the model's representational capacity is limited to the distribution of the text-speech paired data, resulting in poor robustness for unlabeled speech-only data (e.g., different styles, environmental sounds, channels, or qualities).

In our previous three works: 1) We identified issues of information redundancy and dimensional explosion in existing semantic coding methods\cite{qiang2024minimally}. 2) We proposed a semantic coding method called ``Contrastive Token-Acoustic Pretraining (CTAP)\cite{qiang2024learning}." 3) Building on CTAP, we developed a minimally-supervised high-fidelity speech synthesis method\cite{qiang2024high}. Building on these foundational works, we now introduce VQ-CTAP, which uses the cross-modal aligned sequence transcoder to bring text and speech into a joint multimodal space, facilitating the connection between text and speech at the frame level. Our contributions in this paper are as follows:

1. We propose a cross-modal sequence representation learning paradigm and introduce the VQ-CTAP model trained on 900 hours of speech-text pairs and 20,000 hours of speech-only data. We design a stepping optimization strategy to ensure effective model convergence by gradually injecting and adjusting the influence of various loss components.

2. We propose a semantic-transfer-wise paralinguistic consistency loss, which uses unlabeled data to enhance representation capabilities while improving the semantic-paralinguistic decoupling ability of the representations. This loss function allows the model to better generalize to unseen data and capture the nuances of paralinguistic information.

3. We propose a sequence-aware semantic connector, enabling the pre-trained VQ-CTAP to be directly used for TTS task without fine-tuning. Additionally, the pre-trained VQ-CTAP can be directly applied to VC and ASR tasks without fine-tuning or adding additional structures, exhibiting a plug-and-play capability.

4. We achieve high-compression speech coding by generating discrete embedding from a single codebook at a rate of 25Hz from 24kHz input waveforms, which is a 960-fold reduction in the sampling rate. Furthermore, it can effectively improve the prediction stability when applied to autoregressive structures in TTS.
\section{Related work}

\subsection{Speech Representaion}

Self-supervised speech representation learning with deep neural networks emerge two training schemes, the autoregressive models \cite{baevskivq, chung2019unsupervised} and bidirectional models \cite{ling2020decoar, liu2020mockingjay, baevski2020wav2vec, Hsu2021HuBERTSS, chung2021w2vbert, chen2022wavlm}. The autoregressive models learn to predict the future discrete representations of audio segments based on past observations. The bidirectional models mainly adopt masked language modeling (MLM)\cite{devlin2019bert}, which recovers the masked part of the discrete inputs using the unmasked context. These descrete inputs can be derived from the audio segments\cite{ling2020decoar, baevski2020wav2vec}. Following the MLM training scheme for speech representation learning , HuBERT\cite{Hsu2021HuBERTSS} proposed to learn the masked prediction of hidden units generated by vanilla acoustic unit discovery systems. BEST-RQ\cite{chiu2022selfsupervised} learn a model to predict the masked labels of the speech signals generated with a random-projection quantizer. WavLM\cite{chen2022wavlm} jointly learn masked speech prediction and denoising in pre-training to solve full-stack downstream speech tasks.

\subsection{Contrastive Learning}

Contrastive learning has proven to be an effective method for cross-modal modeling. It works by differentiating a target sample (positive) from distractor samples (negatives) based on an anchor representation. The objective is to maximize the similarity between the anchor and positive samples while minimizing the similarity between the anchor and negative samples. This approach has been widely applied in the field of computer vision, with notable examples such as Open AI's CLIP\cite{radford2021learning}, Florence\cite{yuan2021florence}, and ALIGN\cite{jia2021scaling}.

In the audio field, CLIP-based models have also been developed, including Wav2CLIP\cite{wu2022wav2clip}, AudioCLIP\cite{guzhov2022audioclip}, and CLAP\cite{elizalde2023clap}. These models focus on extracting global descriptive information from audio, with the primary goal of improving the performance of downstream audio classification tasks. The downstream tasks applied in the experimental parts of these studies are primarily speech classification tasks, such as sound event classification, instrument classification, acoustic scene classification, emotion recognition, keyword spotting, vocal sound classification, and speaker counting. Since the global features extracted by these methods lose temporal information, they cannot be converted back into frame-level acoustic features. As a result, they are unsuitable for tasks like TTS, VC, and ASR\cite{mohamed2022self}.

\subsection{TTS, VC, and ASR tasks}
The TTS task refers to the modelling of unequal length sequences between text to speech. As shown in Figure \ref{fig:tts-vc-asr}, minimally-supervised TTS refers to splitting the TTS task into two tasks (text-to-semantic and semantic-to-acoustic) by combining speech intermediate representations. As deep learning advances, TTS technology has made significant progress. Traditional TTS methods have achieved satisfactory results\cite{wang2017tacotron,arik2017deep, li2019neural,ren2019fastspeech, kim2020glow, elias2021parallel}. The introduction and subsequent evolution of generative pre-trained transformers (GPT) models\cite{radford2018improving, brown2020language}, have significantly heightened interest in the development of large-scale TTS systems. These TTS systems can be broadly divided into two categories: 1) autoregressive frameworks \cite{borsos2023audiolm, wang2023neural,zhang2023speak,kharitonov2023speak} and 2) non-autoregressive frameworks \cite{levkovitch2022zero, shen2023naturalspeech, le2023voicebox}. 

The VC task aims to modify the voice of a source speaker to match a target style, including aspects such as speaker identity, prosody, and emotion, while preserving the original linguistic content. VC techniques can be broadly classified based on their approach to content disentanglement into two categories: text-based VC and text-free VC. Text-based VC often employs an ASR model to derive PPGs as a representation of content\cite{sun2016phonetic, liu2021any, zhang2021transfer}. In contrast, text-free VC approaches aim to bypass the need for text annotations by directly learning content representations\cite{wu2020vqvc+, chen2021again, li2023freevc}. 

The ASR is the process of converting spoken language into text. This process encompasses several key components, including feature extraction, acoustic modeling, and language modeling. ASR is moving towards speech representation learning to make full use of the large-scale unsupervised speech datasets\cite{chiu2022selfsupervised, Hsu2021HuBERTSS, chung2021w2vbert, chen2022wavlm}. The MLM\cite{ling2020decoar, baevski2020wav2vec} based self-supervised learning machanism learn discrete masked speech prediction to derive the augmented speech representation,
ASR plays as the downstream speech task benefits from the strong speech representation from large-scale datas based self-supervised learning method\cite{ling2020decoar, liu2020mockingjay,chung2021w2vbert}

\begin{figure*}[t]
 \centering
 \includegraphics[width=\linewidth]{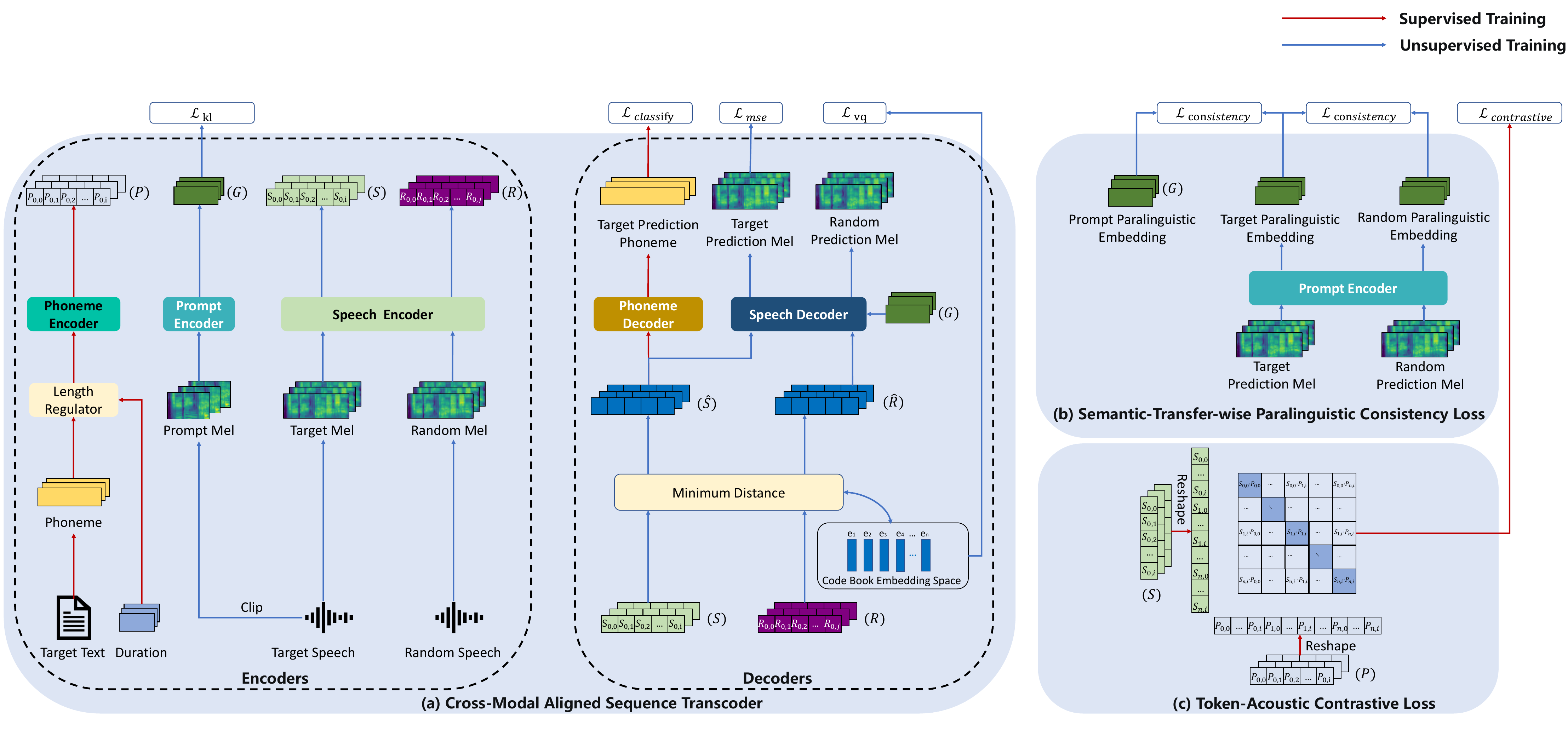}
 \caption{The main body of VQ-CTAP is the cross-modal aligned sequence transcoder, which jointly trains three encoders and two decoders. It extracts phoneme embedding $(P)$ from the target text, target speech embedding $(S)$ and prompt paralinguistic embedding $(G)$ from the target speech, as well as random speech embedding $(R)$ from the random speech. $(P)$ and $(S)$ are used to construct contrastive token-acoustic pre-training, which learns frame-level (dis)similarity between a batch of speech and text pairs. Additionally, two decoders are adapted for downstream tasks such as TTS, ASR, and VC. To enable the semantic-paralinguistic decoupling ability of the representation, unlabeled random speech is used to calculate Semantic-Transfer-wise Paralinguistic Consistency Loss.}
 \label{fig:VQCTAP}
\end{figure*}

\section{Method}
As illustrated in Figure \ref{fig:VQCTAP}(a), the main body of VQ-CTAP is the cross-modal aligned sequence transcoder, which comprises five main components: a speech encoder, a phoneme encoder, a prompt encoder, a phoneme decoder, and a speech decoder. A length regulator is employed to address the mismatch between the lengths of phoneme sequences and speech sequences. The input consists of text-speech pairs and unlabeled random speech from outside the target domain. The phoneme and speech representations are concatenated in a joint multimodal space. As shown in Figure \ref{fig:VQCTAP}(c), a contrastive learning approach is used to learn this space by capturing the frame-level (dis)similarity between speech and phoneme pairs. The speech encoder produces discrete embedding at a rate of 25Hz from the 24kHz input waveform, which is a 960-fold reduction in the sampling rate. Vector quantization further removes paralinguistic information irrelevant to semantics from the speech representations. The two decoders adapt the learned representations for downstream tasks such as TTS, VC, and ASR. The method supports both supervised and unsupervised training. To enable domain-enhanced representation learning, semantic-transfer-wise paralinguistic consistency loss is introduced, computed using the unlabeled random speech from outside the target domain, as shown in Figure \ref{fig:VQCTAP}(b). The pre-trained VQ-CTAP can be directly applied to TTS, VC, and ASR tasks without fine-tuning, exhibiting a plug-and-play capability.

\subsection{Cross-Modal Aligned Sequence Transcoder}

{\bf Encoders:}
As illustrated in Figure \ref{fig:VQCTAP} (a), $S_{in}$ denotes the input batch of target speech data, $S_{in} \in \mathbb{R}^{B \times T_s \times D_s}$, where $B$ is the batch size, $T_s$ is the number of time frames, and $D_s$ is the number of spectral components (mel-spectrogram bands). 
$\tilde{P}_{in}$ denotes the input batch of phoneme data, $\tilde{P}_{in} \in \mathbb{R}^{B \times T_p \times D_p}$, where $T_p$ is the number of phoneme codes and $D_p$ is the dimensionality of the phoneme codes. During training, the ground-truth duration is used to extend the length of the phoneme sequences to $T_s$. During inference, the corresponding predicted duration is used. After the length regulator, the phoneme sequences become $P_{in} \in \mathbb{R}^{B \times T_s \times D_p}$, having the same length as the target speech sequences $S_{in}$. The unlabeled random speech $R_{in}$ from outside the target domain is randomly selected, where $R_{in} \in \mathbb{R}^{B \times T_r \times D_s}$. The batch size $B$ and the number of spectral components $D_s$ are consistent with those of $S_{in}$, while typically $T_r \neq T_s$.
The $P_{in}$, $S_{in}$ and $R_{in}$ are then passed through a phoneme encoder and a speech encoder, respectively:

\begin{equation}
\begin{aligned}
P &= \mathrm{PhonemeEncoder}(P_{in});\\
S &= \mathrm{SpeechEncoder}(S_{in});\\
R &= \mathrm{SpeechEncoder}(R_{in})
\end{aligned}
\end{equation}

where ${P} \in \mathbb{R}^{B \times T_s/4 \times d}$ are the phoneme representations, ${S} \in \mathbb{R}^{B \times T_s/4 \times d}$ are the target speech representations, and ${R} \in \mathbb{R}^{B \times T_r/4 \times d}$ are the random speech representations. The two encoders compress the lengths of the phoneme and speech representations by a factor of 4, respectively. We bring ${S}$, ${R}$, and ${P}$ into a joint multimodal space of dimension $d$.

The prompt speech $G_{in}$ is randomly clipped with a window of length 3 seconds at each step as the input to the prompt encoder:

\begin{equation}
\begin{aligned}
 {G} = \mathrm{PromptEncoder}(G_{in})
\end{aligned}
\end{equation}

$G$ denotes the prompt paralinguistic embedding, where $G$ is a fixed-length global vector. $G \in \mathbb{R}^{B \times D_g}$, where $D_g$ is the number of dimensions. The prompt encoder is a VAE-based model \cite{qiang2023improving} that extracts paralinguistic information to address the one-to-many problem in TTS and VC tasks, such as timbre, style, and prosody, from the prompt speech. In Equation (3), $D_{KL}$ refers to the KL loss, $\mathcal{N}(\cdot)$ represents a Gaussian distribution, and $({\hat{\mu}}, {\hat{\sigma}})$ denotes the $(mean, variance)$ of the prompt paralinguistic latent space distribution. To address the KL collapse problem\cite{qiang2022style}, a margin $\Delta$ is introduced to limit the minimum value of the KL loss as shown: 

\begin{equation}
\begin{aligned}
 \mathcal{L}_{kl} = max(0, D_{KL}[\mathcal{N}({\hat{\mu}},{\hat{\sigma}}^2)||\mathcal{N}(0, I)]-\Delta)
\end{aligned}
\end{equation}

{\bf Decoders:}
As shown in Figure \ref{fig:VQCTAP} (a), a discrete codebook component is added to the network for vector quantization processing, which further removes semantic-irrelevant paralinguistic information from the speech representation. In deep learning, a codebook refers to a collection of representative vectors or codes used to quantize data, typically in tasks like vector quantization or feature encoding. It helps reduce dimensionality, improve model efficiency, and enhance performance by mapping high-dimensional inputs to a finite set of meaningful representations\cite{van2017neural}. $S$ and $R$ are compared with all vectors in the codebook, and the codebook vectors with the closest Euclidean distance are used as the vector-quantized speech representation $\hat{S}$ and $\hat{R}$. The vector quantization loss consists of two parts: the commitment loss (which makes the speech representation conform to its closest codebook vector as much as possible) and the codebook loss (which makes the selected codebook vector as close as possible to the speech representation). $sg[\cdot]$ represents ``stop gradient", and $\boldsymbol e$ is the codebook vector.

\begin{equation}
 \mathcal{L}_{vq} = ||S - sg[\boldsymbol e]||^2_2+||sg[S] - \boldsymbol e||^2_2
 \label{L_vq}
\end{equation}

To get faster convergence speed, exponential moving averages\cite{van2017neural} is used instead of codebook loss.

To utilize the pretrained speech representation for downstream ASR task, $\hat{S}$ is fed into a phoneme decoder to predict the phoneme sequence:
\begin{equation}
\begin{aligned}
P_{s} &= \mathrm{PhonemeDecoder}(\hat{S})
\end{aligned}
\end{equation}

The predicted phoneme sequence $P_{s}$ is then compared with the input phoneme sequence $P_{in}$ using a cross-entropy loss:

\begin{equation}
\begin{aligned}
\mathcal{L}_{classify} = CrossEntropy(P_{in}, P_{s})
\end{aligned}
\end{equation}

To enable the pretrained speech representation to be used for downstream TTS and VC tasks, as well as to construct inputs for the subsequent semantic-transfer-wise paralinguistic consistency loss calculation, both $\hat{S}$ and $\hat{R}$ are fed into the same speech decoder to predict the mel-spectrogram. Since $\hat{S}$ and $\hat{R}$ only contain semantic information, $G$ is provided as input to supply paralinguistic information:

\begin{equation}
\begin{aligned}
 S_{s} &= \mathrm{SpeechDecoder}(\hat{S}, G);\\
 R_{r} &= \mathrm{SpeechDecoder}(\hat{R}, G)
\end{aligned}
\end{equation}

It is important to note that although $\hat{S}$ and $\hat{R}$ contain different semantic information, the input $G$ is the same. Therefore, it is expected that the generated $S_{s}$ and $R_{r}$ will have identical paralinguistic information (timbre, prosody, emotion, etc.). Since $G$ is extracted from $S_{in}$, $S_{s}$ is expected to be the same as $S_{in}$. A mean squared error (MSE) loss is used to compare the predicted mel-spectrogram $S_{s}$ with the ground-truth mel-spectrogram $S_{in}$:

\begin{equation}
\begin{aligned}
\mathcal{L}_{mse} = MSE(S_{in}, S_{s})
\end{aligned}
\end{equation}

\subsection{Semantic-Transfer-Wise Paralinguistic Consistency Loss}

Most existing methods use paired tuples of ground truth and synthesized features to calculate the cycle consistency loss\cite{xue2021cycle, an2021improving, joo2020effective, qiang2023improving}. However, due to factors such as ``teacher forcing" and MSE loss, these two features are nearly identical during the training phase, resulting in suboptimal performance. During training, the combination of paralinguistic embedding and transferred semantic embedding (e.g., $G$ from speaker 1 and $\hat{R}$ from speaker 2) as input poses an unseen problem, since there is no ground truth to compute the reconstruction loss. Existing cross-modal fine-grained (frame-level) pre-training methods that aim to explicitly introduce text information as supervision rely on expensive annotated text-speech pairs. In this case, the model's representational capacity is limited to the distribution of the text-speech paired data, resulting in poor robustness for unlabeled data (e.g., different styles, environmental sounds, channels, or qualities).

As shown in Figure \ref{fig:VQCTAP} (b), a semantic-transfer-wise paralinguistic consistency loss is proposed. Unlabeled random speech $R_{in}$ from outside the target domain is randomly selected. The extracted representations $\hat{S}$ and $\hat{R}$ from the target speech and random speech, respectively, are used as inputs, and the speech decoder is computed twice in each training step with identical paralinguistic embedding $G$. We expect the generated $S_{s}$ and $R_{r}$ to have different semantic information but the same paralinguistic information (timbre, prosody, emotion, etc.). $S_{s}$ and $R_{r}$ are then passed through the prompt encoder to extract the target paralinguistic embedding $G_{s}$ and random paralinguistic embedding $G_{r}$, respectively.

We construct two paralinguistic consistency losses: ($G$ \& $G_{s}$) and ($G_{s}$ \& $G_{r}$). This approach helps the model learn unseen semantic-paralinguistic combinations during the training phase, introduces unlabeled data, and enhances the model's robustness. The paralinguistic consistency loss is calculated using the Gram matrix\cite{ma2018neural}, which captures the local statistics of the audio signal in the frequency and time domains:

\begin{equation}
\begin{aligned}
 \mathcal{L}_{consistency1} = \frac{1}{n^2} \sum(G^T*G - G_{s}^T*G_{s})^2;\\
\mathcal{L}_{consistency2} = \frac{1}{n^2} \sum(G_{s}^T*G_{s} - G_{r}^T*G_{r})^2;\\
\end{aligned}
\end{equation}

\begin{algorithm*}
\caption{Training process of the VQ-CTAP}\label{alg:training}
\begin{algorithmic}
\STATE Init $step$, $kl\_start$, $kl\_end$, $consistency\_start$, $consistency\_end$, $kl\_upper$, $consistency\_upper$
\FOR{each $step$}
    \STATE Update model with $\mathcal{L}_{vq} + \mathcal{L}_{classify} + \mathcal{L}_{mse} + \mathcal{L}_{contrastive}$
    \IF{$step > kl\_start$}
        \STATE $beat = kl\_upper \times \min(1, \frac{(step - kl\_start)}{(kl\_end - kl\_start)})$
        \STATE Update model with $\mathcal{L}_{kl} \times {beat}$
    \ENDIF
    \IF{$step > consistency\_start$}
        \STATE $beat = consistency\_upper \times \min(1, \frac{(step - consistency\_start)}{(consistency\_end - consistency\_start)})$
        \STATE Update model with $\mathcal{L}_{consistency} \times {beat}$
    \ENDIF
\ENDFOR
\end{algorithmic}
\end{algorithm*}

\subsection{Token-Acoustic Contrastive Loss}

As shown in Figure \ref{fig:VQCTAP} (c), to extract frame-level representations, $S$ and $P$ within a batch are reshaped into 2D matrices $S_{re}$ and $P_{re}$, where $S_{re}$ and $P_{re} \in \mathbb{R}^{(B * T_s) \times d}$. This approach is beneficial for contrastive learning, as it increases the number of sample pairs per step, thereby enhancing the learning process.

With $S_{re}$ and $P_{re}$ now comparable, we can measure their similarity:

\begin{equation}
    C = \tau*({S}_{re} \cdot {P}_{re}^\top)
\end{equation}

{\noindent}where $\tau$ is a temperature parameter used to scale the range of logits. The resulting similarity matrix $C \in \mathbb{R}^{(B * T_s) \times (B * T_s)}$ contains $(B * T_s)$ correct pairs along the diagonal and $(B * T_s)^2-(B * T_s)$ incorrect pairs in the off-diagonal elements. As the extracted intermediate representation includes contextual information, only the current frame corresponds to a positive sample.

The contrastive loss is calculated:

\begin{equation}
     \mathcal{L}_{contrastive}= 0.5 * (\ell_{speech}(C) + \ell_{phoneme}(C))
\end{equation}

{\noindent}where 

\begin{equation}
     \ell_{(speech \text{ or } phoneme)} = \frac{1}{B * T_s}\sum_{i=0}^{B * T_s} \log diag (softmax(C))
\end{equation}


{\noindent}along the speech and phoneme axes, respectively. This symmetric cross-entropy loss ($\mathcal{L}_{contrastive}$) is computed over the similarity matrix to jointly train the speech encoder and the phoneme encoder, enabling the model to learn meaningful representations from both modalities simultaneously.

\subsection{Stepping Optimization Strategy}

A stepping optimization strategy is designed to ensure effective model convergence by gradually injecting and adjusting the influence of various loss components, as shown in Algorithm \ref{alg:training}. The training process involves the following losses: $\mathcal{L}_{kl}$, $\mathcal{L}_{vq}$, $\mathcal{L}_{classify}$, $\mathcal{L}_{mse}$, $\mathcal{L}_{contrastive}$, and $\mathcal{L}_{consistency}$. The variable $step$ represents the current training step.

Initially, the model is trained using $\mathcal{L}_{vq}$, $\mathcal{L}_{mse}$, $\mathcal{L}_{contrastive}$, and $\mathcal{L}_{classify}$. Each of these losses has a corresponding weight, with the weights for $\mathcal{L}_{mse}$, $\mathcal{L}_{vq}$, and $\mathcal{L}_{classify}$ being equal, while the weight for $\mathcal{L}_{contrastive}$ is significantly smaller than the other losses.
When the $step$ exceeds the specified starting step for $\mathcal{L}_{kl}$, $\mathcal{L}_{kl}$ is added to the training process. The weight for $\mathcal{L}_{kl}$ increases gradually as the training progresses. Once the $step$ surpasses the specified ending step, the weight for $\mathcal{L}_{kl}$ is fixed at $kl\_upper$.
Similarly, when the $step$ exceeds the specified starting step for $\mathcal{L}_{consistency}$, $\mathcal{L}_{consistency}$ is incorporated into the training process. The weight for $\mathcal{L}_{consistency}$ also increases gradually during training. Once the $step$ exceeds the specified ending step, the weight for $\mathcal{L}_{consistency}$ is fixed at $consistency\_upper$.

Algorithm \ref{alg:training} outlines the step-wise inclusion of different losses and their corresponding weight adjustments. This optimization strategy aims to facilitate effective model convergence by gradually introducing and adjusting the influence of various loss components throughout the training process.

\begin{figure*}
 \centering
 \includegraphics[width=0.7\linewidth]{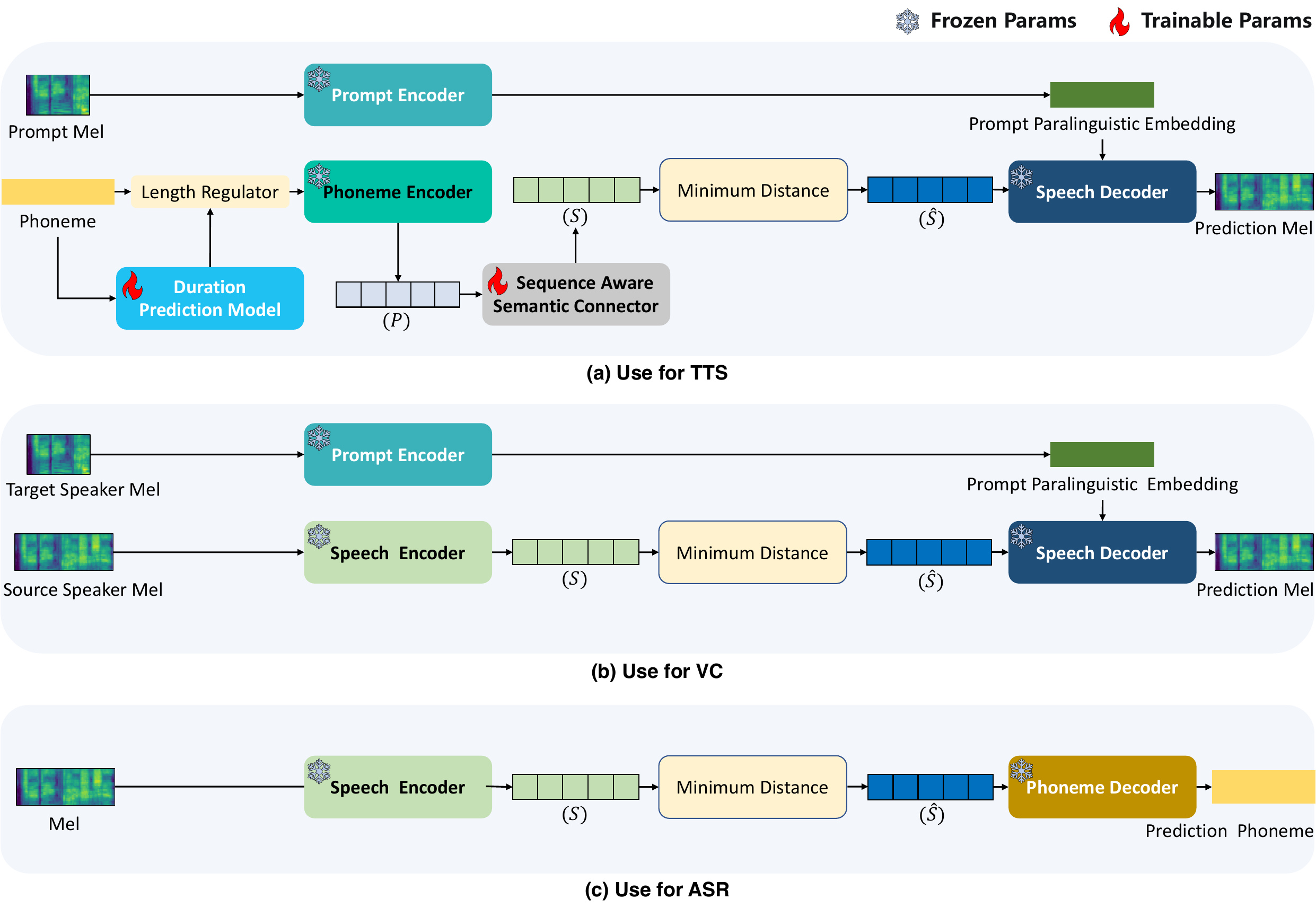}
 \caption{The pre-trained VQ-CTAP is used for downstream TTS, VC and ASR tasks.}
 \label{fig:application}
\end{figure*}

\begin{figure*}
 \centering
 \includegraphics[width=0.7\linewidth]{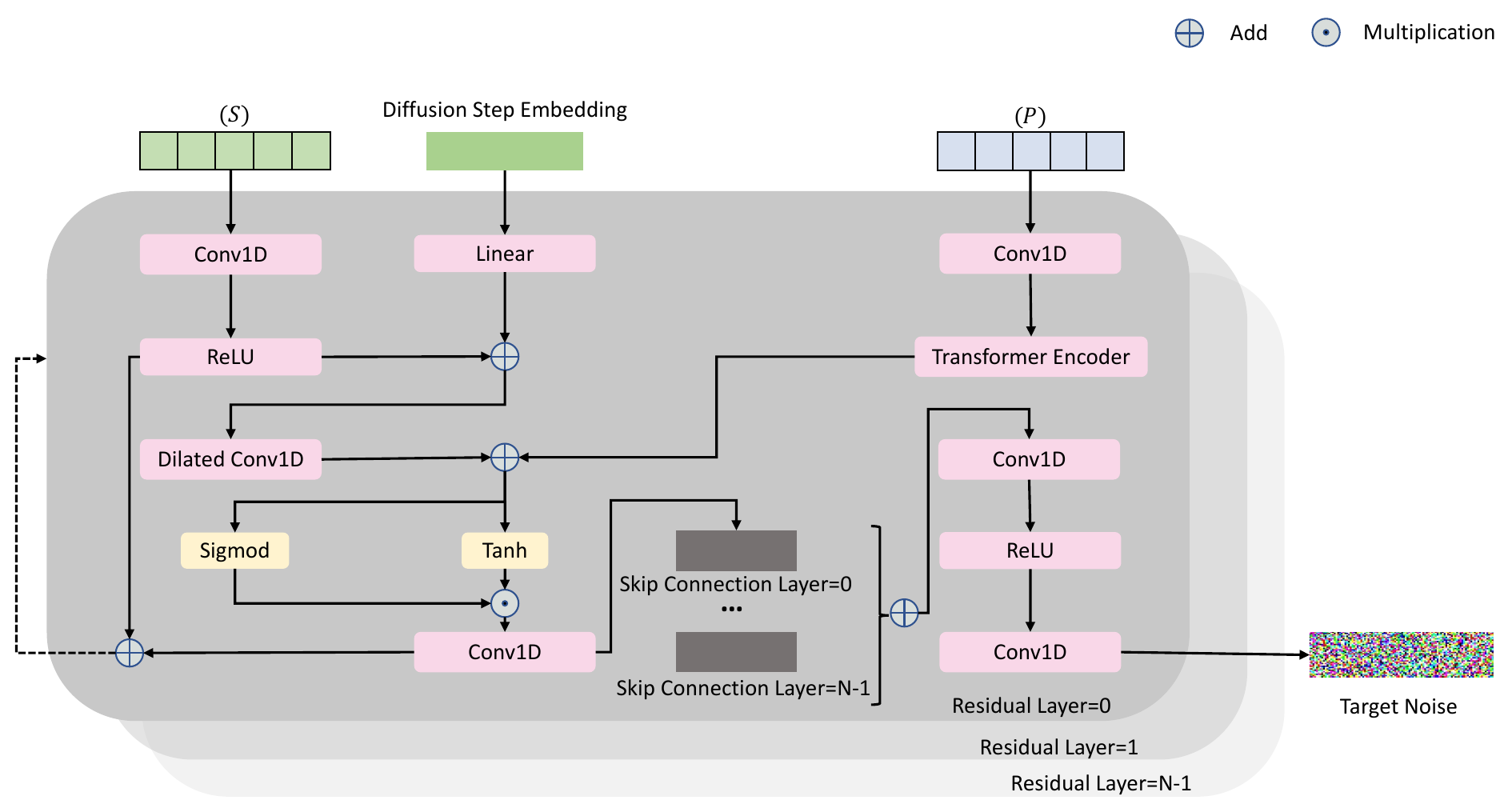}
 \caption{The architecture of sequence-aware semantic connector}
 \label{fig:connector}
\end{figure*}

\section{Plug-and-Play for Downstream Tasks}

\subsection{TTS Pipeline}

VQ-CTAP can be applied to the task of minimally supervised TTS. Minimally supervised TTS refers to splitting the TTS task into two sub-tasks: text2semantic and semantic2acoustic, by combining speech intermediate representations. As shown in Figure \ref{fig:application}(a), the weights of the pre-trained phoneme encoder, prompt encoder, speech encoder, and speech decoder are frozen. A diffusion-based duration model is utilized to predict the duration corresponding to each phoneme. The pre-trained phoneme encoder is used to extract phoneme embedding, while the pre-trained prompt encoder is used to extract prompt paralinguistic embedding. 
We find that phoneme representations $P$ and speech representations $S$, due to the frame-level token-acoustic contrastive loss, are very close in Euclidean space. VQ-CTAP can be used for TTS without fine-tuning or a connector, but the quality is suboptimal due to the gap between $P$ and $S$. Therefore, we proposed the Sequence-Aware Semantic Connector to achieve high-quality TTS at minimal cost.
The presence of $\mathcal{L}_{consistency}$ and the model's inherent semantic-acoustic decoupling design naturally align $P$ with $S$, making it easier for the sequence-aware semantic connector to predict the semantic-aligned speech embedding $S$. The predicted speech embedding $S$ is quantized using a codebook to obtain quantized speech embedding $\hat{S}$, along with the prompt paralinguistic embedding, are fed into the pre-trained speech decoder to generate the predicted mel-spectrogram.

\subsubsection{Sequence-Aware Semantic Connector}

The sequence-aware semantic connector calculation is shown in Algorithms \ref{alg:connector_training} and \ref{alg:connector_sampling}. The model uses $q(data), S_0, P,$ and $t$ to represent data distribution, speech embedding, phoneme embedding, and diffusion step respectively. A notable feature of the model is its ability to perform closed-form sampling of $S_t$ at any timestep $t$ using $\bar{\alpha}_t$ and $\alpha_t$.  The non-autoregressive network $\epsilon_{\theta}$ predicts $\epsilon$ from $S_t$, $t$, and $P$. The training objective is to minimize the unweighted variant of the ELBO \cite{ho2020denoising}, as shown in line 6 of Algorithm \ref{alg:connector_training}. The sampling process, detailed in Algorithm \ref{alg:connector_sampling}, begins by sampling $S_T \sim \mathcal{N}(0, I)$, followed by iteratively sampling $S_{t-1}\sim p_{\theta}(S_{t-1}|S_t)$ for $t=T, T-1,\cdots,1$. The output $S_0$ represents the sampled data.

\begin{algorithm*}
\caption{Training of Sequence-Aware Semantic Connector} \label{alg:connector_training}
\begin{algorithmic}
\REPEAT
  \STATE $S_0, P \sim q(data)$
  \STATE $t \sim \mathrm{Uniform}(\{1, \dotsc, T\})$, $\varepsilon \sim \mathcal{N}(0, I)$, $\bar{\alpha}_t = \prod_{i=1}^t\alpha_i$
  \STATE Take gradient descent step on
  \STATE \hspace{0.5cm} $\nabla _\theta \left\| \varepsilon - \varepsilon_\theta((\sqrt{\bar\alpha_t} S_0 + \sqrt{1-\bar\alpha_t}\varepsilon), t, P) \right\|^2$
\UNTIL{converged}
\end{algorithmic}
\end{algorithm*}

\begin{algorithm*}
\caption{Sampling of Sequence-Aware Semantic Connector} \label{alg:connector_sampling}
\begin{algorithmic}
\STATE $S_T \sim \mathcal{N}(0, I)$
\FOR{$t = T, \dotsc, 1$}
  \STATE $\mu_{\theta}(S_t, t, P) = \frac{1}{\sqrt{\alpha_t}} \left( S_t - \frac{1-\alpha_t}{\sqrt{1-\bar\alpha_t}} \varepsilon_\theta(S_t, t, P) \right)$
  \STATE $\sigma_{\theta}(S_t, t, P) = \sqrt{\frac{1-\bar{\alpha}_{t-1}}{1-\bar{\alpha}_t}(1-\alpha_t)}$
  \STATE $S_{t-1} = \mu_{\theta} + \sigma_{\theta} \odot \psi$; $\psi \sim \mathcal{N}(0, I)$ if $t > 1$, else $\psi = 0$
\ENDFOR
\STATE \textbf{return} $S_0$
\end{algorithmic}
\end{algorithm*}

Figure \ref{fig:connector} illustrates the architecture of the sequence-aware semantic connector, which employs a bidirectional dilated convolution structure with $N$ residual layers organized into $m$ blocks, each containing $n = \frac{N}{m}$ layers. Within each block, the dilation is doubled at each layer. Skip connections from all residual layers are summed, akin to the approach used in WaveNet \cite{oord2016wavenet}. The dashed arrow indicates that the output of the current layer serves as the input to the next residual layer. The model incorporates phoneme embedding as conditional information, which is processed by a transformer encoder and added as bias terms to the dilated convolution in each residual layer. The diffusion step embedding is broadcast over length and added to the input of each residual layer.

\subsection{VC Pipeline}
VQ-CTAP can be directly applied to the VC task without fine-tuning, as illustrated in Figure \ref{fig:application}(b). In this process, the weights of the prompt encoder, speech encoder, and speech decoder remain frozen. The pre-trained speech encoder extracts the speech embedding of the source speaker, while the pre-trained prompt encoder extracts the prompt paralinguistic embedding of the target speaker. The speech embedding $S$ is then quantized using a codebook to obtain the quantized speech embedding $\hat{S}$. Subsequently, $\hat{S}$ and prompt paralinguistic embedding is fed together into the pre-trained speech decoder. The predicted mel spectrogram from the speech decoder combines the semantic content of the source speaker with the timbre of the target speaker.

\subsection{ASR Pipeline}
VQ-CTAP can be directly applied to the ASR task without fine-tuning, as illustrated in Figure \ref{fig:application}(c). In this approach, the weights of the speech encoder and phoneme decoder are frozen. The pre-trained speech encoder is utilized to extract speech embedding $S$. $S$ is quantized using a codebook to obtain the quantized speech embedding $\hat{S}$. Finally, $\hat{S}$ is fed into the pre-trained phoneme decoder, which predicts the corresponding phoneme sequence.

\section{Experiments Procedures}

\subsection{Model Details}
In the experiment, we use a speech encoder with a structure similar to the Whisper encoder \cite{radford2023robust}, but reduce the dimension to $d$ through a linear layer.
The speech encoder consists of 2 convolutional layers, a GELU activation function, 6 transformer layers, and a linear layer. The 2 convolutional layers compress the length to a quarter of the original shape.
The phoneme encoder is composed of a convolutional layer, a ReLU activation function, 4 transformer layers, and a linear layer.
The outputs of the speech encoder and phoneme encoder are layer-normalized separately.
The prompt encoder is a VAE-based model, consisting of 6 convolutional layers and an SE-ResNet\cite{hu2018squeeze} block.
The codebook has 8192 embedding entries, with each embedding having a dimension of 256.
The speech decoder comprises 6 transformer layers, 5 convolutional layers, 2 transposed convolutional layers (which increase the length by a factor of 4 to restore the original shape), 6 Tanh activation functions, and a linear layer.
The phoneme decoder consists of 6 transformer layers, 2 transposed convolutional layers (which increase the length by a factor of 4 to restore the original length), 2 Tanh activation functions, and a linear layer.
Both encoders and decoders have a hidden dimension of 256.
The sequence-aware semantic connector consists of 30 residual layers, 64 residual channels, a kernel size of 3, and a dilation cycle of $[1, 2, \cdots, 512]$. The diffusion step is set to 50, with a linearly spaced schedule $\beta_{t} \in [1 \times 10^{-4}, 0.05]$ ($T=200$).  
For the stepping optimization strategy, the parameters are as follows: $kl\_start = 1e4$, $kl\_end = 2e4$, $consistency\_start = 2e4$, $consistency\_end = 3e4$, $kl\_upper = 1e-5$, and $consistency\_upper = 1e-5$.  
The model is trained using 8 NVIDIA TESLA A800 80GB GPUs, with a batch size of 64 per GPU. Adam\cite{Kingma2014AdamAM} is used as the optimizer, with an initial learning rate of 2e-4. The number of sample pairs per step ranges from 4,000 to 32,000 during the training process.

\subsection{Datasets}

For the labeled text-speech paired data, we integrate our internal dataset with the AISHELL-3 dataset \cite{shi2020aishell} and the LibriTTS dataset \cite{zen2019libritts}, resulting in a combined total of 900 hours of recordings from 3000 speakers.

Regarding the unlabeled speech-only data, we trawl 20,000 hours of speech from the internet, primarily sourced from audiobooks, podcasts, and online videos. However, this portion of the data lacks speaker information statistics. The collected audio underwent various preprocessing steps, including Voice Activity Detection (VAD) segmentation, signal-to-noise ratio (SNR) detection, and human voice detection.

All speech waveforms are sampled at 24kHz and converted to 40-band mel spectrograms with a frame size of 960 and a hop size of 240. From an information compression perspective, 40-dimensional features are more suitable for our task.

\subsection{Compared Method and Tasks}

To demonstrate the model's capability for downstream frame-level tasks, we evaluated its performance on TTS, VC, and ASR tasks. We compare our proposed model with five other models, including self-supervised learning methods: {\bf Codec}\cite{Defossez2022HighFN}, {\bf Hubert}\cite{Hsu2021HuBERTSS}, {\bf Wav2Vec2.0}\cite{baevski2020wav2vec}, and supervised learning methods: the encoder of the ASR model {\bf Whisper}\cite{radford2023robust}, and the speech representation model {\bf CTAP} \cite{qiang2024learning} based on Contrastive Phoneme-Speech Pretraining. In this paper, the representations extracted from these models are collectively referred to as semantic coding, as shown in Tabel\ref{tab:description}. In all models, the embedding values from the codebook are used to replace discrete codes, resulting in dense representation of the inputs and outputs. To ensure fairness, when constructing the TTS, VC, and ASR systems, only the intermediate semantic coding (representations extracted by these models) is different, the remaining structures are completely identical consistent with Tabe\ref{fig:application} and trained using the same dataset for fairness. Our model frameworks are aligned with the state-of-the-art methods: SoVITS\footnote{https://github.com/svc-develop-team/so-vits-svc} for VC and SPEAR-TTS\cite{kharitonov2023speak} for minimally-supervised TTS. All comparison models were trained from scratch using the same dataset. The memory sizes of the speech encoders for VQ-CTAP, Wav2Vec 2.0, Encodec, HuBERT, and Whisper are 105MB, 310MB, 57MB, 308MB, and 244MB, respectively. While VQ-CTAP has a denser structure compared to other models, it has fewer parameters than HuBERT and Wav2Vec 2.0. Moreover, VQ-CTAP uniquely supports plug-and-play downstream tasks, a feature not offered by the other models.
Many disentangled VQ codecs are multi-codebook models\cite{ju2024naturalspeech, zhang2023speechtokenizer, defossez2024moshi}, while VQ-CTAP is a single-codebook model. In single-codebook scenarios, VQ-CTAP performs significantly better than these models, which is why we did not choose them as comparison methods.

Additionally, we conduct the following ablation experiments: {\bf VQ-CTAP w/o PhonemeDecoder} does not include $\mathcal{L}_{classify}$ during training. {\bf VQ-CTAP w/o Compression} removes the length compression and restoration operations from the encoders/decoders, {\bf VQ-CTAP w/o Contrastive} does not include $\mathcal{L}_{contrastive}$ during training. {\bf VQ-CTAP w/o Consistency} does not include the semantic-transfer-wise paralinguistic consistency loss during training. {\bf VQ-CTAP w/o Unsupervised} only uses text-speech paired data during training. {\bf VQ-CTAP w/o Stepping} does not use the stepping optimization strategy during training, and all loss weights are the same. The {\bf VQ-CTAP (Ground Truth)} structure uses intermediate representations extracted from ground truth speech for TTS.

To validate the effectiveness of the sequence-aware semantic connector, we separately train a two-stage TTS system {\bf VQ-CTAP + Separate} based on VQ-CTAP representations as a control model.

\begin{figure*}
 \centering
 \includegraphics[width=0.75\linewidth]{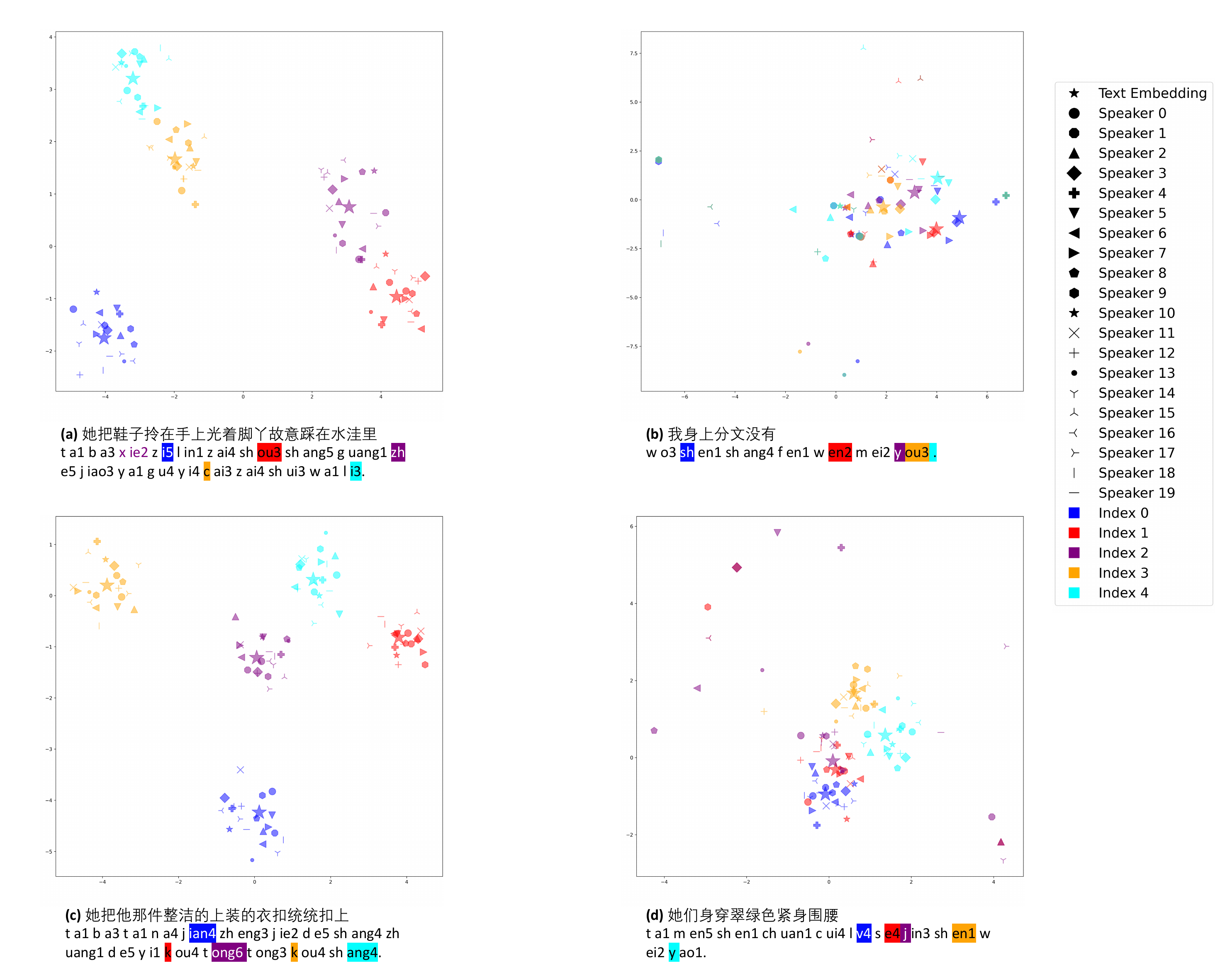}
 \caption{t-SNE plot of phoneme/speech embedding for 20 speakers. The $\bigstar$ represents $P$, and the other shapes represent $S$ for different speakers, with different colors indicating the positions of corresponding $P$ and $S$. For the red and orange phonemes ``k", although the phonemes are the same and the positions are different, the corresponding $P$ and $S$ are not entangled.}
 \label{fig:tsne_phoneme_embedding}
\end{figure*}

\begin{figure*}
 \centering
 \includegraphics[width=0.75\linewidth]{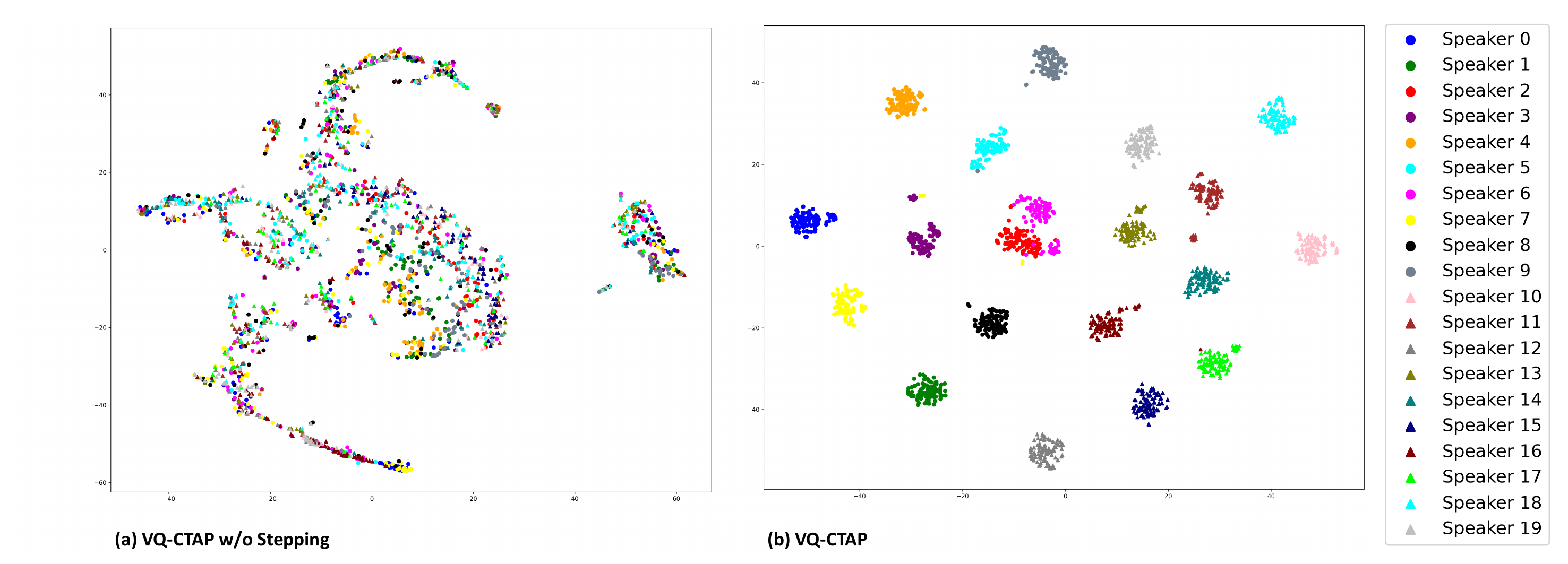}
 \caption{t-SNE plot of prompt paralinguistic embedding for 20 speakers with 100 sentences each}
 \label{fig:tsne_prompt_embedding}
\end{figure*}

\subsection{Evaluation Metrics}

For the TTS and VC tasks, we conduct all subjective tests using 25 paid human evaluators to assess speech quality, prosody, and timbre similarity. The evaluators have been standardized on a unified 1-5 scoring system, with reference examples provided for each score level. Before participating in the official evaluation, they must pass a test using these standard examples to ensure their understanding and consistency.

The evaluation metrics used in the assessment include prosody measurement, which involves mean square error for pitch ({\bf MSEP}) to assess prosody similarity against ground-truth speech; word error rate ({\bf WER}) (2000 sentences), utilizing an ASR model to transcribe the generated speech; and mean opinion score (MOS), to validate the speech quality ({\bf Q-MOS}) and the similarity of expected speaker prosody ({\bf P-MOS}) and speaker timbre ({\bf S-MOS}) between the source speech and the synthesized speech. In the ASR task, we compare the phoneme accuracy ({\bf ACC}) (2000 sentences) of each method.

\begin{table}
 \caption{Comparison of Compression}
 \label{tab:compression}
 \centering
\resizebox{\linewidth}{!}{ 
\begin{tabular}{lccc}
\hline
Model       & Speech Sampling Rate & Coding Sample Rate & Compression           \\ \hline
Encodec\cite{Defossez2022HighFN}     & 24 kHz               & 75 Hz              & 320 $\times$          \\ \hline
SoundStream\cite{zeghidour2021soundstream} & 24 kHz               & 75 Hz              & 320 $\times$          \\ \hline
Hubert\cite{Hsu2021HuBERTSS}      & 16 kHz               & 50 Hz              & 320 $\times$          \\ \hline
Wav2vec2.0\cite{baevski2020wav2vec}  & 16 kHz               & 49 Hz              & 326 $\times$          \\ \hline
CTAP\cite{qiang2024learning}        & 24 kHz               & 100 Hz             & 240 $\times$          \\ \hline
VQ-CTAP     & 24 kHz               & 25 Hz              & \textbf{960 $\times$} \\ \hline
\end{tabular}
}
\end{table}

\begin{table}[t]
 \caption{Description of Tasks}
 \label{tab:description}
 \centering
\resizebox{\linewidth}{!}{ 
\begin{tabular}{l|l}
\hline
{TTS} & $ text \rightarrow  semantic \ prediction \ model  \rightarrow$ \\
                     & $ intermediate \ representation + prompt \ embedding \rightarrow$  \\
                     & $  acoustic \ prediction \ model \rightarrow \ mel$ \\ \hline
{VC}  & $ speech \rightarrow \ semantic \ extractor \rightarrow $ \\
                     & $ intermediate \ representation +  prompt \ embedding \rightarrow$ \\
                     & $ acoustic \ prediction \ model \rightarrow \ mel$ \\ \hline
{ASR} & $ speech \rightarrow \ semantic \ extractor \rightarrow$ \\
                     & $ intermediate \ representation \rightarrow $ \\
                     & $  phoneme \ prediction \ model \rightarrow \ phoneme$ \\ \hline
\end{tabular}
}
\end{table}

\begin{table*}[t]
 \caption{Results of Minimally-Supervised TTS, VC, and ASR}
 \label{tab:compare}
 \centering
\resizebox{\linewidth}{!}{ 
\begin{tabular}{l|ccccc|cccc|c}
\hline
Model                & \multicolumn{5}{c|}{TTS Task}                                                                            & \multicolumn{4}{c|}{VC Task}                                                          & ASR Task         \\ \cline{2-11} 
                     & MSEP$\downarrow$ & WER$\downarrow$  & P-MOS$\uparrow$      & S-MOS$\uparrow$      & Q-MOS$\uparrow$      & WER$\downarrow$  & P-MOS$\uparrow$      & S-MOS$\uparrow$      & Q-MOS$\uparrow$      & ACC$\uparrow$    \\ \hline
Encodec\cite{Defossez2022HighFN}                & 95.3             & 7.0              & 3.64 ± 0.07          & 3.75 ± 0.07          & 3.68 ± 0.14          & 4.8              & 4.01 ± 0.02          & 2.88 ± 0.11          & 3.55 ± 0.05          & \textbackslash{} \\ \hline
Wav2Vec2.0\cite{baevski2020wav2vec}           & 107.2            & 5.1              & 3.79 ± 0.10          & 3.71 ± 0.09          & 3.80 ± 0.12          & 4.6              & 3.98± 0.09           & 3.41 ± 0.09          & 3.47 ± 0.05          & 92.04            \\ \hline
Hubert\cite{Hsu2021HuBERTSS}               & 103.2            & 5.2              & 3.82 ± 0.03          & 3.70 ± 0.02          & 3.81 ± 0.02          & 4.4              & 3.93 ± 0.12          & 3.45 ± 0.05          & \textbf{3.90 ± 0.07} & \textbf{96.28}   \\ \hline
Whisper Encoder\cite{radford2023robust}      & \textbackslash{} & \textbackslash{} & \textbackslash{}     & \textbackslash{}     & \textbackslash{}     & 6.1              & 3.79 ± 0.04          & 3.19 ± 0.02          & 3.57 ± 0.04          & 96.13            \\ \hline
CTAP\cite{qiang2024learning}                 & 99.1             & 4.4              & 3.89± 0.06           & 3.91 ± 0.08          & \textbf{3.88 ± 0.10} & 4.3              & 3.99 ± 0.13          & 3.78 ± 0.19          & 3.87 ± 0.10          & 91.70            \\ \hline
VQ-CTAP(Groud Truth) & 63.9             & 2.0              & 4.43± 0.02           & 4.53 ± 0.05          & 4.19 ± 0.07          & \textbackslash{} & \textbackslash{}     & \textbackslash{}     & \textbackslash{}     & \textbackslash{} \\ \hline
VQ-CTAP              & \textbf{92.9}    & 4.1              & \textbf{3.99 ± 0.09} & \textbf{4.03 ± 0.01} & 3.79 ± 0.03          & \textbf{3.9}     & \textbf{4.10 ± 0.11} & \textbf{4.12 ± 0.03} & 3.81 ± 0.13          & 93.73            \\ \hline
\end{tabular}
}
\end{table*}

\begin{figure*}[b]
 \centering
 \includegraphics[width=0.8\linewidth]{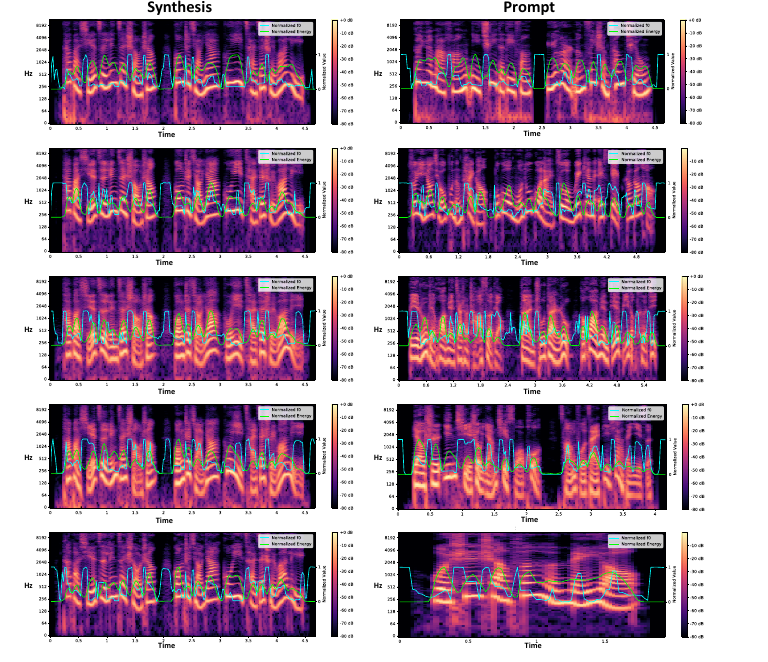}
 \caption{The Spectrograms, F0, and Energy of synthesized speech (the same semantic coding combining different prompts). The bottom row is the ground-truth. The synthesized speech and the prompt speech exhibit consistency in the numerical range and variation trends of spectrogram, F0, and energy.}
 
 \label{fig:mel_f0_energy}
\end{figure*}

\begin{table*}[t]
 \caption{Results of Ablation Studies}
 \label{tab:ablation}
 \centering
\resizebox{\linewidth}{!}{ 
\begin{tabular}{l|ccccc|cccc|c}
\hline
Model                      & \multicolumn{5}{c|}{TTS Task}                                                                           & \multicolumn{4}{c|}{VC Task}                                                         & ASR Task       \\ \cline{2-11} 
                           & MSEP$\downarrow$ & WER$\downarrow$ & P-MOS$\uparrow$      & S-MOS$\uparrow$      & Q-MOS$\uparrow$      & WER$\downarrow$ & P-MOS$\uparrow$      & S-MOS$\uparrow$      & Q-MOS$\uparrow$      & ACC$\uparrow$  \\ \hline
VQ-CTAP w/o PhonemeDecoder & 95.9             & 4.9             & 3.97 ± 0.12          & 3.94 ± 0.09          & 3.82 ± 0.02          & 4.3             & 4.04 ± 0.03          & 3.99 ± 0.13          & 3.81 ± 0.14          & 88.12          \\ \hline
VQ-CTAP w/o Compression    & 93.7             & \textbf{3.8}    & \textbf{4.01 ± 0.03} & 4.02 ± 0.02          & 3.80± 0.10           & 3.9             & 4.09 ± 0.07          & 4.01 ± 0.03          & 3.88 ± 0.04          & \textbf{94.01} \\ \hline
VQ-CTAP w/o Contrastive    & 99.9             & 5.0             & 3.79 ± 0.09          & 3.61 ± 0.11          & 3.75 ± 0.08          & 5.3             & 3.76 ± 0.14          & 3.59 ± 0.02          & 3.69 ± 0.09          & 91.73          \\ \hline
VQ-CTAP w/o Consistency    & 98.7             & 4.4             & 3.90 ± 0.02          & 3.98 ± 0.14          & 3.79 ± 0.13          & 4.2             & 4.00 ± 0.08          & 3.99 ± 0.12          & 3.79 ± 0.08          & 92.98          \\ \hline
VQ-CTAP w/o Unsupervised   & 100.9            & 4.7             & 3.81 ± 0.15          & 3.87 ± 0.04          & \textbf{3.88 ± 0.13} & 4.9             & 3.93 ± 0.04          & 3.92.± 0.14          & \textbf{3.90 ± 0.16} & 91.88          \\ \hline
VQ-CTAP w/o Stepping       & 118.5            & 5.9             & 3.65 ± 0.21          & 3.62 ± 0.19          & 3.65 ± 0.16          & 6.0             & 3.67 ± 0.13          & 3.49 ± 0.12          & 3.66 ± 0.19          & 87.23          \\ \hline
VQ-CTAP                    & \textbf{92.9}    & 4.1             & 3.99 ± 0.09          & \textbf{4.03 ± 0.01} & 3.79 ± 0.03          & \textbf{3.9}    & \textbf{4.10 ± 0.11} & \textbf{4.12 ± 0.03} & 3.81 ± 0.13          & 93.73          \\ \hline
\end{tabular}
}
\end{table*}

\section{Results}

\subsection{Compression and Decoupling Ability}
Table \ref{tab:compression} compares the compression ratios of VQ-CTAP and mainstream speech representation models. It shows that the 960x compression ratio of VQ-CTAP is 3 times that of Encodec, SoundStream, and Hubert, while the experiments prove that VQ-CTAP maintains high performance(Table \ref{tab:compare}) even at high compression rates.

To intuitively demonstrate that VQ-CTAP's semantic coding is paralinguistic-independent, we use prompts from 20 speakers to synthesize 4 text, with each synthesis having the same duration to ensure that the text corresponding to each frame of audio is the same across different speakers. We then use VQ-CTAP's speech encoder to extract the speech representation $S$ and the phoneme encoder to extract the text's phoneme representation $P$. We expect these two representations to be aligned at the frame level, and $S$ should have removed the speaker information. Using t-SNE, we compress each frame's $P$ and $S$ to 2-dimensional vectors and show their distribution in Figure \ref{fig:tsne_phoneme_embedding}. 
As shown in Figures \ref{fig:tsne_phoneme_embedding}(a) and (c), both the Euclidean distance among the $S$ of different speakers and the corresponding $P$ at each frame are very close, proving that VQ-CTAP’s speech encoder extracts the speech representation (semantic coding) while removing paralinguistic information and aligning it with the text modality. Notably, for the red and orange phonemes ``k" in Figure \ref{fig:tsne_phoneme_embedding}(c), although the phonemes are the same and the positions are different, the corresponding $P$ and $S$ are not entangled. This indicates that VQ-CTAP's representation is highly context-dependent, which is in line with our motivation for using the token-acoustic contrastive loss. The clustering effect in the short sentences (b) and (d) of Figure \ref{fig:tsne_phoneme_embedding} is not as good as that in the long sentences (a) and (c), also confirming VQ-CTAP's characteristic of context modeling.

We randomly select 2000 utterances from 20 speakers and use VQ-CTAP's prompt encoder to extract the prompt paralinguistic embedding. We compress it to a 2-dimensional vector based on the t-SNE as shown in Figure \ref{fig:tsne_prompt_embedding}. The scattered distribution can be seen in Figure \ref{fig:tsne_prompt_embedding}(a), proving the prompt encoder cannot model timbre without the stepping optimization strategy. In contrast, the prompt paralinguistic embedding extracted by VQ-CTAP has clear boundaries between different speakers, demonstrating the prompt encoder's timbre modeling capability. (Except for {VQ-CTAP w/o Stepping}, the other ablation models also have this capability.)

VQ-CTAP has better efficiency similar to other contrastive learning-based methods, and it significantly reduces computational complexity compared to other pre-training schemes, i.e., it only predicts frame-level aligned phoneme and speech. The large number of frames included in each step of our calculations (batch size multiplied by sequence length) is advantageous for contrastive learning.

\subsection{Evaluation of TTS Task}
Table \ref{tab:compare} presents the results of the TTS task. {\bf VQ-CTAP} achieves the best results in 4 terms of MSEP, WER, P-MOS, and S-MOS. Incorporating the decoder structure and reconstruction loss during training helps extract intermediate representations that are more suitable for frame-level generation tasks. Multiple systems achieve similar results in terms of prosody similarity P-MOS since they employ the same diffusion-based duration model. However, {\bf VQ-CTAP} achieves the best prosody similarity P-MOS due to its enhanced semantic-paralinguistic consistency, enabled by modules such as the Semantic-Transfer-wise Paralinguistic Consistency Loss. {\bf CTAP} and {\bf VQ-CTAP} address the issue of speaker information redundancy in traditional intermediate representations through frame-level contrastive learning. {\bf VQ-CTAP} exhibits less voice leakage, leading to the best results in terms of voice similarity S-MOS. Furthermore, the pre-trained phoneme encoder alleviates the difficulty of predicting intermediate representations (semantic embeddings) from text in two-stage TTS models and can be directly used as an effective predictor. In terms of speech quality Q-MOS, {\bf Codec}, {\bf Hubert}, {\bf Wav2Vec2.0}, and {\bf VQ-CTAP} suffer from information loss due to their discrete processing, while {\bf CTAP} achieves the best Q-MOS results with its continuous representations that are not subject to discretization.

\subsection{Evaluation of VC Task}
The results of the voice conversion (VC) task are also presented in Table \ref{tab:compare}. Similar to the TTS task, {\bf VQ-CTAP} achieves the best results in terms of WER, P-MOS, and S-MOS. {\bf VQ-CTAP} extracts semantic coding from the source speaker's mel-spectrogram with better decoupling ability for paralinguistic information, and its ability to model the paralinguistic embedding from the target speaker's mel-spectrogram is also stronger. The experiments reveal that {\bf Hubert} has a higher information redundancy issue compared to other models, leading to the retention of more paralinguistic information in the semantic encoding, which helps it achieve the best results in terms of Q-MOS. To point out in the ablation study (Table \ref{tab:ablation}), {\bf CTAP w/o Unsupervised} (without using unsupervised low-quality data) 
achieve similar performance to {\bf Hubert} in terms of Q-MOS.

Figure \ref{fig:mel_f0_energy}  shows the spectrograms, F0, and energy of the VC output speech generated by VQ-CTAP using the same semantic coding combined with different prompts. As shown, the synthesized speech exhibits strong alignment with the prompt speech in terms of spectrogram across multiple frequency domains, F0, and energy. Benefit from the characteristic of decoupling paralinguistic information of the speech encoder in VQ-CTAP and the ability to extract paralinguistic information of the prompt encoder.

\subsection{Evaluation of ASR Task}
Table \ref{tab:compare} shows that {\bf Hubert} achieves the best results in ASR, but is not suitable for TTS and VC tasks, because of higher information redundancy. It's worth noting that in the ablation study (Table \ref{tab:ablation}), {\bf CTAP w/o Compression} (without length compression) achieves results close to {\bf Hubert} in terms of ACC.

\subsection{Ablation Studies}
As shown in Table \ref{tab:ablation}, the ablation study aims to analyze the contribution of the different components in VQ-CTAP. The largest negative impact of the \textbf{VQ-CTAP w/o PhonemeDecoder} model is on the ACC of the ASR task. The phoneme classification loss of the PhonemeDecoder structure is naturally well-suited for the ASR task. Since contrastive learning has already introduced text supervision, its role is similar to that of the phoneme decoder, so removing the phoneme decoder has little impact on the other results.
The \textbf{VQ-CTAP w/o Compression} excels in WER and P-MOS for TTS task and in ACC for ASR task. However, these improvements are relatively minor compared to \textbf{VQ-CTAP}, which maintains excellent reconstruction ability even at high compression rates. Higher compression rates facilitate the separation of paralinguistic information. For autoregressive architectures, such as language models, higher compression rates facilitate easier convergence and enhance stability.
We have experimented with compressing HuBERT and wav2vec 2.0 by 2x or 4x, which resulted in significant loss of content information and performance degradation.
The \textbf{VQ-CTAP w/o Contrastive} model has the most significant negative impact on all metrics, as the token-acoustic contrastive loss is the source of cross-modal modeling capability. \textbf{VQ-CTAP w/o Contrastive} has degraded to an approximate self-supervised representation learning model, lacking the ability to decouple semantics and paralinguistic information.
The \textbf{VQ-CTAP w/o Consistency} model has a decrease in the P-MOS and S-MOS for the TTS and VC tasks. The semantic-transfer-wise paralinguistic consistency loss allows the model to better generalize to unseen data and capture the nuances of paralinguistic information.
The \textbf{VQ-CTAP w/o Unsupervised} model has a significant decline in the MSEP, P-MOS, and S-MOS, as the unlabeled data can enhance the representation capability and improve the semantic-paralinguistic decoupling of the representations. Notably, there is an improvement in the Q-MOS metric, due to the removal of the low-quality unlabeled data.
The \textbf{VQ-CTAP w/o Stepping} model has a very large negative impact on all metrics. Without the stepping optimization strategy during training, the model's KL loss and consistency loss are difficult to converge, and the classify loss and MSE loss values increase, leading to a significant decline in the model's capability.
The experiments show that \textbf{VQ-CTAP} has advantages in downstream speech processing tasks, maintaining P-MOS and S-MOS above 3.99 for the TTS and VC tasks.

Table \ref{tab:connector} verifies the effectiveness of the sequence-aware semantic connector, demonstrating that the method of directly using pre-trained modules for downstream tasks is effective. We separately trained a two-stage TTS model \textbf{VQ-CTAP + Separate}, using the semantic coding extracted by VQ-CTAP, with the text2semantic module and semantic2acoustic module trained separately. The results show that the performance of \textbf{VQ-CTAP} and \textbf{VQ-CTAP + Separate} are very close, indicating that this simple and effective method of using frozen pre-trained modules and adding a connector model is effective.

\subsection{Limitations}
Regarding dataset selection, our method utilizes a small amount of labeled text-speech pair data alongside a large volume of unlabeled speech-only data. Since the model relies on frame-level alignment information, the quality of the labeled data must be as high as possible. Therefore, we selected the widely recognized open-source datasets: AISHELL3\cite{shi2020aishell} for Chinese and LibriTTS\cite{zen2019libritts} for English. In contrast, the unlabeled data does not require frame-level alignment during training, so we used 20,000 hours of data scraped from the Internet.

Concerning potential biases in the evaluation methodology, existing methods validate semantic representation capabilities using ASR tasks. However, in ASR tasks, the decoupling of paralinguistic information is not strongly correlated with ACC. For example, prosody and emotion can affect recognition results. In the future, we aim to explore better evaluation metrics for semantic representation capabilities.

\begin{table}
 \caption{Frozen pre-training modules comparing independently trained models on the TTS task.}
 \label{tab:connector}
 \centering
\resizebox{\linewidth}{!}{ 
\begin{tabular}{lccccc}
\hline
Model              & MSEP$\downarrow$          & WER$\downarrow$          & P-MOS$\uparrow$                & S-MOS$\uparrow$                & Q-MOS$\uparrow$                \\ \hline
VQ-CTAP + Separate & \textbf{91.4} & 4.2          & 3.95 ± 0.08          & \textbf{4.07 ± 0.04} & \textbf{3.82 ± 0.05} \\ \hline
VQ-CTAP            & 92.9          & \textbf{4.1} & \textbf{3.99 ± 0.09} & 4.03 ± 0.01          & 3.79 ± 0.03          \\ \hline
\end{tabular}
}
\end{table}

\section{Conclusions and future work}
This paper proposes a cross-modal fine-grained sequence representation learning method VQ-CTAP, which has a plug-and-play capability for downstream tasks and can be directly applied to TTS, VC, and ASR tasks without fine-tuning. The introduction of unsupervised speech-only data has enhanced generalization capabilities. The key contributions include the cross-modal aligned sequence transcoder, the semantic-transfer-wise paralinguistic consistency loss, the token-acoustic contrastive loss, the sequence-aware semantic connector, and the stepping optimization strategy, all of which effectively enhance the model's representation capability and semantic-paralinguistic decoupling. Additionally, VQ-CTAP achieves a high compression ratio of 960x for speech coding while maintaining high performance. In future work, we will explore incorporating other modalities, such as images and videos, to achieve multi-modal sequence representation learning.

\bibliographystyle{IEEEtran}
\bibliography{trans}

\begin{thebibliography}{10}
\providecommand{\url}[1]{#1}
\csname url@samestyle\endcsname
\providecommand{\newblock}{\relax}
\providecommand{\bibinfo}[2]{#2}
\providecommand{\BIBentrySTDinterwordspacing}{\spaceskip=0pt\relax}
\providecommand{\BIBentryALTinterwordstretchfactor}{4}
\providecommand{\BIBentryALTinterwordspacing}{\spaceskip=\fontdimen2\font plus
\BIBentryALTinterwordstretchfactor\fontdimen3\font minus \fontdimen4\font\relax}
\providecommand{\BIBforeignlanguage}[2]{{%
\expandafter\ifx\csname l@#1\endcsname\relax
\typeout{** WARNING: IEEEtran.bst: No hyphenation pattern has been}%
\typeout{** loaded for the language `#1'. Using the pattern for}%
\typeout{** the default language instead.}%
\else
\language=\csname l@#1\endcsname
\fi
#2}}
\providecommand{\BIBdecl}{\relax}
\BIBdecl

\bibitem{radford2021learning}
A.~Radford, J.~W. Kim, C.~Hallacy, A.~Ramesh, G.~Goh, S.~Agarwal, G.~Sastry, A.~Askell, P.~Mishkin, J.~Clark \emph{et~al.}, ``Learning transferable visual models from natural language supervision,'' in \emph{International conference on machine learning}.\hskip 1em plus 0.5em minus 0.4em\relax PMLR, 2021, pp. 8748--8763.

\bibitem{yuan2021florence}
L.~Yuan, D.~Chen, Y.-L. Chen, N.~Codella, X.~Dai, J.~Gao, H.~Hu, X.~Huang, B.~Li, C.~Li \emph{et~al.}, ``Florence: A new foundation model for computer vision,'' \emph{arXiv preprint arXiv:2111.11432}, 2021.

\bibitem{jia2021scaling}
C.~Jia, Y.~Yang, Y.~Xia, Y.-T. Chen, Z.~Parekh, H.~Pham, Q.~Le, Y.-H. Sung, Z.~Li, and T.~Duerig, ``Scaling up visual and vision-language representation learning with noisy text supervision,'' in \emph{International conference on machine learning}.\hskip 1em plus 0.5em minus 0.4em\relax PMLR, 2021, pp. 4904--4916.

\bibitem{wu2022wav2clip}
H.-H. Wu, P.~Seetharaman, K.~Kumar, and J.~P. Bello, ``Wav2clip: Learning robust audio representations from clip,'' in \emph{ICASSP 2022-2022 IEEE International Conference on Acoustics, Speech and Signal Processing (ICASSP)}.\hskip 1em plus 0.5em minus 0.4em\relax IEEE, 2022, pp. 4563--4567.

\bibitem{guzhov2022audioclip}
A.~Guzhov, F.~Raue, J.~Hees, and A.~Dengel, ``Audioclip: Extending clip to image, text and audio,'' in \emph{ICASSP 2022-2022 IEEE International Conference on Acoustics, Speech and Signal Processing (ICASSP)}.\hskip 1em plus 0.5em minus 0.4em\relax IEEE, 2022, pp. 976--980.

\bibitem{elizalde2023clap}
B.~Elizalde, S.~Deshmukh, M.~Al~Ismail, and H.~Wang, ``Clap learning audio concepts from natural language supervision,'' in \emph{ICASSP 2023-2023 IEEE International Conference on Acoustics, Speech and Signal Processing (ICASSP)}.\hskip 1em plus 0.5em minus 0.4em\relax IEEE, 2023, pp. 1--5.

\bibitem{girdhar2023imagebind}
R.~Girdhar, A.~El-Nouby, Z.~Liu, M.~Singh, K.~V. Alwala, A.~Joulin, and I.~Misra, ``Imagebind: One embedding space to bind them all,'' in \emph{Proceedings of the IEEE/CVF Conference on Computer Vision and Pattern Recognition}, 2023, pp. 15\,180--15\,190.

\bibitem{zhu2023languagebind}
B.~Zhu, B.~Lin, M.~Ning, Y.~Yan, J.~Cui, W.~HongFa, Y.~Pang, W.~Jiang, J.~Zhang, Z.~Li \emph{et~al.}, ``Languagebind: Extending video-language pretraining to n-modality by language-based semantic alignment,'' in \emph{The Twelfth International Conference on Learning Representations}.

\bibitem{kharitonov2023speak}
E.~Kharitonov, D.~Vincent, Z.~Borsos, R.~Marinier, S.~Girgin, O.~Pietquin, M.~Sharifi, M.~Tagliasacchi, and N.~Zeghidour, ``Speak, read and prompt: High-fidelity text-to-speech with minimal supervision,'' \emph{Transactions of the Association for Computational Linguistics}, vol.~11, pp. 1703--1718, 2023.

\bibitem{baevski2020wav2vec}
A.~Baevski, Y.~Zhou, A.~Mohamed, and M.~Auli, ``wav2vec 2.0: A framework for self-supervised learning of speech representations,'' in \emph{NeurIPS}, 2020.

\bibitem{sadhu2021wav2vec}
S.~Sadhu, D.~He, C.-W. Huang, S.~H. Mallidi, M.~Wu, A.~Rastrow, A.~Stolcke, J.~Droppo, and R.~Maas, ``wav2vec-c: A self-supervised model for speech representation learning,'' in \emph{Proc. Interspeech 2021}, 2021, pp. 711--715.

\bibitem{baevski2019vq}
A.~Baevski, S.~Schneider, and M.~Auli, ``vq-wav2vec: Self-supervised learning of discrete speech representations,'' \emph{arXiv preprint arXiv:1910.05453}, 2019.

\bibitem{Hsu2021HuBERTSS}
W.-N. Hsu, B.~Bolte, Y.-H.~H. Tsai, K.~Lakhotia, R.~Salakhutdinov, and A.~Mohamed, ``Hubert: Self-supervised speech representation learning by masked prediction of hidden units,'' \emph{IEEE/ACM Transactions on Audio, Speech, and Language Processing}, vol.~29, pp. 3451--3460, 2021.

\bibitem{chung2021w2v}
Y.-A. Chung, Y.~Zhang, W.~Han, C.-C. Chiu, J.~Qin, R.~Pang, and Y.~Wu, ``W2v-bert: Combining contrastive learning and masked language modeling for self-supervised speech pre-training,'' in \emph{2021 IEEE Automatic Speech Recognition and Understanding Workshop (ASRU)}.\hskip 1em plus 0.5em minus 0.4em\relax IEEE, 2021, pp. 244--250.

\bibitem{sun2016phonetic}
L.~Sun, K.~Li, H.~Wang, S.~Kang, and H.~Meng, ``Phonetic posteriorgrams for many-to-one voice conversion without parallel data training,'' in \emph{2016 IEEE International Conference on Multimedia and Expo (ICME)}.\hskip 1em plus 0.5em minus 0.4em\relax IEEE, 2016, pp. 1--6.

\bibitem{qiang2024minimally}
C.~Qiang, H.~Li, H.~Ni, H.~Qu, R.~Fu, T.~Wang, L.~Wang, and J.~Dang, ``Minimally-supervised speech synthesis with conditional diffusion model and language model: A comparative study of semantic coding,'' in \emph{ICASSP 2024-2024 IEEE International Conference on Acoustics, Speech and Signal Processing (ICASSP)}.\hskip 1em plus 0.5em minus 0.4em\relax IEEE, 2024, pp. 10\,186--10\,190.

\bibitem{qiang2024learning}
C.~Qiang, H.~Li, Y.~Tian, R.~Fu, T.~Wang, L.~Wang, and J.~Dang, ``Learning speech representation from contrastive token-acoustic pretraining,'' in \emph{ICASSP 2024-2024 IEEE International Conference on Acoustics, Speech and Signal Processing (ICASSP)}.\hskip 1em plus 0.5em minus 0.4em\relax IEEE, 2024, pp. 10\,196--10\,200.

\bibitem{qiang2024high}
C.~Qiang, H.~Li, Y.~Tian, Y.~Zhao, Y.~Zhang, L.~Wang, and J.~Dang, ``High-fidelity speech synthesis with minimal supervision: All using diffusion models,'' in \emph{ICASSP 2024-2024 IEEE International Conference on Acoustics, Speech and Signal Processing (ICASSP)}.\hskip 1em plus 0.5em minus 0.4em\relax IEEE, 2024, pp. 10\,781--10\,785.

\bibitem{baevskivq}
A.~Baevski, S.~Schneider, and M.~Auli, ``vq-wav2vec: Self-supervised learning of discrete speech representations,'' in \emph{International Conference on Learning Representations}.

\bibitem{chung2019unsupervised}
Y.-A. Chung, W.-N. Hsu, H.~Tang, and J.~Glass, ``An unsupervised autoregressive model for speech representation learning,'' in \emph{Interspeech}, 2019.

\bibitem{ling2020decoar}
S.~Ling and Y.~Liu, ``Decoar 2.0: Deep contextualized acoustic representations with vector quantization,'' \emph{arXiv preprint arXiv:2012.06659}, 2020.

\bibitem{liu2020mockingjay}
A.~T. Liu, S.-w. Yang, P.-H. Chi, P.-c. Hsu, and H.-y. Lee, ``Mockingjay: Unsupervised speech representation learning with deep bidirectional transformer encoders,'' in \emph{ICASSP}, 2020.

\bibitem{chung2021w2vbert}
Y.-A. Chung, Y.~Zhang, W.~Han, C.-C. Chiu, J.~Qin, R.~Pang, and Y.~Wu, ``W2v-bert: Combining contrastive learning and masked language modeling for self-supervised speech pre-training,'' in \emph{2021 IEEE Automatic Speech Recognition and Understanding Workshop (ASRU)}.\hskip 1em plus 0.5em minus 0.4em\relax IEEE, 2021, pp. 244--250.

\bibitem{chen2022wavlm}
S.~Chen, C.~Wang, Z.~Chen, Y.~Wu, S.~Liu, Z.~Chen, J.~Li, N.~Kanda, T.~Yoshioka, X.~Xiao \emph{et~al.}, ``Wavlm: Large-scale self-supervised pre-training for full stack speech processing,'' \emph{IEEE Journal of Selected Topics in Signal Processing}, vol.~16, no.~6, pp. 1505--1518, 2022.

\bibitem{devlin2019bert}
J.~Devlin, M.-W. Chang, K.~Lee, and K.~Toutanova, ``Bert: Pre-training of deep bidirectional transformers for language understanding,'' in \emph{NAACL}, 2019.

\bibitem{chiu2022selfsupervised}
C.-C. Chiu, J.~Qin, Y.~Zhang, J.~Yu, and Y.~Wu, ``Self-supervised learning with random-projection quantizer for speech recognition,'' in \emph{ICML}, 2022.

\bibitem{mohamed2022self}
A.~Mohamed, H.-y. Lee, L.~Borgholt, J.~D. Havtorn, J.~Edin, C.~Igel, K.~Kirchhoff, S.-W. Li, K.~Livescu, L.~Maal{\o}e \emph{et~al.}, ``Self-supervised speech representation learning: A review,'' \emph{IEEE Journal of Selected Topics in Signal Processing}, 2022.

\bibitem{wang2017tacotron}
Y.~Wang, R.~Skerry-Ryan, D.~Stanton, Y.~Wu, R.~J. Weiss, N.~Jaitly, Z.~Yang, Y.~Xiao, Z.~Chen, S.~Bengio \emph{et~al.}, ``Tacotron: Towards end-to-end speech synthesis,'' in \emph{Proc. Interspeech 2017}, 2017, pp. 4006--4010.

\bibitem{arik2017deep}
S.~{\"O}. Ar{\i}k, M.~Chrzanowski, A.~Coates, G.~Diamos, A.~Gibiansky, Y.~Kang, X.~Li, J.~Miller, A.~Ng, J.~Raiman \emph{et~al.}, ``Deep voice: Real-time neural text-to-speech,'' in \emph{International Conference on Machine Learning}.\hskip 1em plus 0.5em minus 0.4em\relax PMLR, 2017, pp. 195--204.

\bibitem{li2019neural}
N.~Li, S.~Liu, Y.~Liu, S.~Zhao, and M.~Liu, ``Neural speech synthesis with transformer network,'' in \emph{Proceedings of the AAAI Conference on Artificial Intelligence}, vol.~33, no.~01, 2019, pp. 6706--6713.

\bibitem{ren2019fastspeech}
Y.~Ren, Y.~Ruan, X.~Tan, T.~Qin, S.~Zhao, Z.~Zhao, and T.-Y. Liu, ``Fastspeech: Fast, robust and controllable text to speech,'' \emph{Advances in Neural Information Processing Systems}, vol.~32, 2019.

\bibitem{kim2020glow}
J.~Kim, S.~Kim, J.~Kong, and S.~Yoon, ``Glow-tts: A generative flow for text-to-speech via monotonic alignment search,'' \emph{Advances in Neural Information Processing Systems}, vol.~33, pp. 8067--8077, 2020.

\bibitem{elias2021parallel}
I.~Elias, H.~Zen, J.~Shen, Y.~Zhang, Y.~Jia, R.~J. Weiss, and Y.~Wu, ``Parallel tacotron: Non-autoregressive and controllable tts,'' in \emph{ICASSP 2021-2021 IEEE International Conference on Acoustics, Speech and Signal Processing (ICASSP)}.\hskip 1em plus 0.5em minus 0.4em\relax IEEE, 2021, pp. 5709--5713.

\bibitem{radford2018improving}
A.~Radford, K.~Narasimhan, T.~Salimans, and I.~Sutskever, ``Improving language understanding by generative pre-training,'' 2018.

\bibitem{brown2020language}
T.~Brown, B.~Mann, N.~Ryder, M.~Subbiah, J.~D. Kaplan, P.~Dhariwal, A.~Neelakantan, P.~Shyam, G.~Sastry, A.~Askell \emph{et~al.}, ``Language models are few-shot learners,'' \emph{Advances in neural information processing systems}, vol.~33, pp. 1877--1901, 2020.

\bibitem{borsos2023audiolm}
Z.~Borsos, R.~Marinier, D.~Vincent, E.~Kharitonov, O.~Pietquin, M.~Sharifi, D.~Roblek, O.~Teboul, D.~Grangier, M.~Tagliasacchi \emph{et~al.}, ``Audiolm: a language modeling approach to audio generation,'' \emph{IEEE/ACM transactions on audio, speech, and language processing}, vol.~31, pp. 2523--2533, 2023.

\bibitem{wang2023neural}
S.~Chen, C.~Wang, Y.~Wu, Z.~Zhang, L.~Zhou, S.~Liu, Z.~Chen, Y.~Liu, H.~Wang, J.~Li, L.~He, S.~Zhao, and F.~Wei, ``Neural codec language models are zero-shot text to speech synthesizers,'' \emph{IEEE Transactions on Audio, Speech and Language Processing}, vol.~33, pp. 705--718, 2025.

\bibitem{zhang2023speak}
Z.~Zhang, L.~Zhou, C.~Wang, S.~Chen, Y.~Wu, S.~Liu, Z.~Chen, Y.~Liu, H.~Wang, J.~Li \emph{et~al.}, ``Speak foreign languages with your own voice: Cross-lingual neural codec language modeling,'' \emph{arXiv preprint arXiv:2303.03926}, 2023.

\bibitem{levkovitch2022zero}
A.~Levkovitch, E.~Nachmani, and L.~Wolf, ``Zero-shot voice conditioning for denoising diffusion tts models,'' in \emph{Proc. Interspeech 2022}, 2022, pp. 2983--2987.

\bibitem{shen2023naturalspeech}
K.~Shen, Z.~Ju, X.~Tan, E.~Liu, Y.~Leng, L.~He, T.~Qin, J.~Bian \emph{et~al.}, ``Naturalspeech 2: Latent diffusion models are natural and zero-shot speech and singing synthesizers,'' in \emph{The Twelfth International Conference on Learning Representations}.

\bibitem{le2023voicebox}
M.~Le, A.~Vyas, B.~Shi, B.~Karrer, L.~Sari, R.~Moritz, M.~Williamson, V.~Manohar, Y.~Adi, J.~Mahadeokar \emph{et~al.}, ``Voicebox: Text-guided multilingual universal speech generation at scale,'' \emph{Advances in neural information processing systems}, vol.~36, pp. 14\,005--14\,034, 2023.

\bibitem{liu2021any}
S.~Liu, Y.~Cao, D.~Wang, X.~Wu, X.~Liu, and H.~Meng, ``Any-to-many voice conversion with location-relative sequence-to-sequence modeling,'' \emph{IEEE/ACM Transactions on Audio, Speech, and Language Processing}, vol.~29, pp. 1717--1728, 2021.

\bibitem{zhang2021transfer}
M.~Zhang, Y.~Zhou, L.~Zhao, and H.~Li, ``Transfer learning from speech synthesis to voice conversion with non-parallel training data,'' \emph{IEEE/ACM Transactions on Audio, Speech, and Language Processing}, vol.~29, pp. 1290--1302, 2021.

\bibitem{wu2020vqvc+}
D.-Y. Wu, Y.-H. Chen, and H.-y. Lee, ``Vqvc+: One-shot voice conversion by vector quantization and u-net architecture,'' in \emph{Proc. Interspeech 2020}, 2020, pp. 4691--4695.

\bibitem{chen2021again}
Y.-H. Chen, D.-Y. Wu, T.-H. Wu, and H.-y. Lee, ``Again-vc: A one-shot voice conversion using activation guidance and adaptive instance normalization,'' in \emph{ICASSP 2021-2021 IEEE International Conference on Acoustics, Speech and Signal Processing (ICASSP)}.\hskip 1em plus 0.5em minus 0.4em\relax IEEE, 2021, pp. 5954--5958.

\bibitem{li2023freevc}
J.~Li, W.~Tu, and L.~Xiao, ``Freevc: Towards high-quality text-free one-shot voice conversion,'' in \emph{ICASSP 2023-2023 IEEE International Conference on Acoustics, Speech and Signal Processing (ICASSP)}.\hskip 1em plus 0.5em minus 0.4em\relax IEEE, 2023, pp. 1--5.

\bibitem{qiang2023improving}
C.~Qiang, P.~Yang, H.~Che, Y.~Zhang, X.~Wang, and Z.~Wang, ``Improving prosody for cross-speaker style transfer by semi-supervised style extractor and hierarchical modeling in speech synthesis,'' in \emph{ICASSP 2023-2023 IEEE International Conference on Acoustics, Speech and Signal Processing (ICASSP)}.\hskip 1em plus 0.5em minus 0.4em\relax IEEE, 2023, pp. 1--5.

\bibitem{qiang2022style}
C.~Qiang, P.~Yang, H.~Che, X.~Wang, and Z.~Wang, ``Style-label-free: Cross-speaker style transfer by quantized vae and speaker-wise normalization in speech synthesis,'' in \emph{2022 13th International Symposium on Chinese Spoken Language Processing (ISCSLP)}, 2022, pp. 61--65.

\bibitem{van2017neural}
A.~Van Den~Oord, O.~Vinyals \emph{et~al.}, ``Neural discrete representation learning,'' \emph{Advances in neural information processing systems}, vol.~30, 2017.

\bibitem{xue2021cycle}
L.~Xue, S.~Pan, L.~He, L.~Xie, and F.~K. Soong, ``Cycle consistent network for end-to-end style transfer tts training,'' \emph{Neural Networks}, vol. 140, pp. 223--236, 2021.

\bibitem{an2021improving}
X.~An, F.~K. Soong, and L.~Xie, ``Improving performance of seen and unseen speech style transfer in end-to-end neural tts,'' in \emph{Proc. Interspeech 2021}, 2021, pp. 4688--4692.

\bibitem{joo2020effective}
Y.-S. Joo, H.~Bae, Y.-I. Kim, H.-Y. Cho, and H.-G. Kang, ``Effective emotion transplantation in an end-to-end text-to-speech system,'' \emph{IEEE Access}, vol.~8, pp. 161\,713--161\,719, 2020.

\bibitem{ma2018neural}
S.~Ma, D.~Mcduff, and Y.~Song, ``Neural tts stylization with adversarial and collaborative games,'' in \emph{International Conference on Learning Representations}, 2018.

\bibitem{ho2020denoising}
J.~Ho, A.~Jain, and P.~Abbeel, ``Denoising diffusion probabilistic models,'' \emph{Advances in neural information processing systems}, vol.~33, pp. 6840--6851, 2020.

\bibitem{oord2016wavenet}
A.~v.~d. Oord, S.~Dieleman, H.~Zen, K.~Simonyan, O.~Vinyals, A.~Graves, N.~Kalchbrenner, A.~Senior, and K.~Kavukcuoglu, ``Wavenet: A generative model for raw audio,'' \emph{arXiv preprint arXiv:1609.03499}, 2016.

\bibitem{radford2023robust}
A.~Radford, J.~W. Kim, T.~Xu, G.~Brockman, C.~McLeavey, and I.~Sutskever, ``Robust speech recognition via large-scale weak supervision,'' in \emph{International Conference on Machine Learning}.\hskip 1em plus 0.5em minus 0.4em\relax PMLR, 2023, pp. 28\,492--28\,518.

\bibitem{hu2018squeeze}
J.~Hu, L.~Shen, and G.~Sun, ``Squeeze-and-excitation networks,'' in \emph{Proceedings of the IEEE conference on computer vision and pattern recognition}, 2018, pp. 7132--7141.

\bibitem{Kingma2014AdamAM}
D.~P. Kingma and J.~Ba, ``Adam: A method for stochastic optimization,'' \emph{CoRR}, vol. abs/1412.6980, 2014.

\bibitem{shi2020aishell}
Y.~Shi, H.~Bu, X.~Xu, S.~Zhang, and M.~Li, ``Aishell-3: A multi-speaker mandarin tts corpus,'' in \emph{Proc. Interspeech 2021}, 2021, pp. 2756--2760.

\bibitem{zen2019libritts}
H.~Zen, V.~Dang, R.~Clark, Y.~Zhang, R.~J. Weiss, Y.~Jia, Z.~Chen, and Y.~Wu, ``Libritts: A corpus derived from librispeech for text-to-speech,'' in \emph{Proc. Interspeech 2019}, 2019, pp. 1526--1530.

\bibitem{Defossez2022HighFN}
A.~D{\'e}fossez, J.~Copet, G.~Synnaeve, and Y.~Adi, ``High fidelity neural audio compression,'' \emph{Transactions on Machine Learning Research}.

\bibitem{ju2024naturalspeech}
Z.~Ju, Y.~Wang, K.~Shen, X.~Tan, D.~Xin, D.~Yang, E.~Liu, Y.~Leng, K.~Song, S.~Tang \emph{et~al.}, ``Naturalspeech 3: Zero-shot speech synthesis with factorized codec and diffusion models,'' in \emph{International Conference on Machine Learning}.\hskip 1em plus 0.5em minus 0.4em\relax PMLR, 2024, pp. 22\,605--22\,623.

\bibitem{zhang2023speechtokenizer}
X.~Zhang, D.~Zhang, S.~Li, Y.~Zhou, and X.~Qiu, ``Speechtokenizer: Unified speech tokenizer for speech language models,'' in \emph{The Twelfth International Conference on Learning Representations}.

\bibitem{defossez2024moshi}
A.~D{\'e}fossez, L.~Mazar{\'e}, M.~Orsini, A.~Royer, P.~P{\'e}rez, H.~J{\'e}gou, E.~Grave, and N.~Zeghidour, ``Moshi: a speech-text foundation model for real-time dialogue,'' \emph{arXiv preprint arXiv:2410.00037}, 2024.

\bibitem{zeghidour2021soundstream}
N.~Zeghidour, A.~Luebs, A.~Omran, J.~Skoglund, and M.~Tagliasacchi, ``Soundstream: An end-to-end neural audio codec,'' \emph{IEEE/ACM Transactions on Audio, Speech, and Language Processing}, vol.~30, pp. 495--507, 2021.

\end{thebibliography}

\newpage

\begin{IEEEbiography}[{\includegraphics[width=1in,height=1.25in,clip,keepaspectratio]{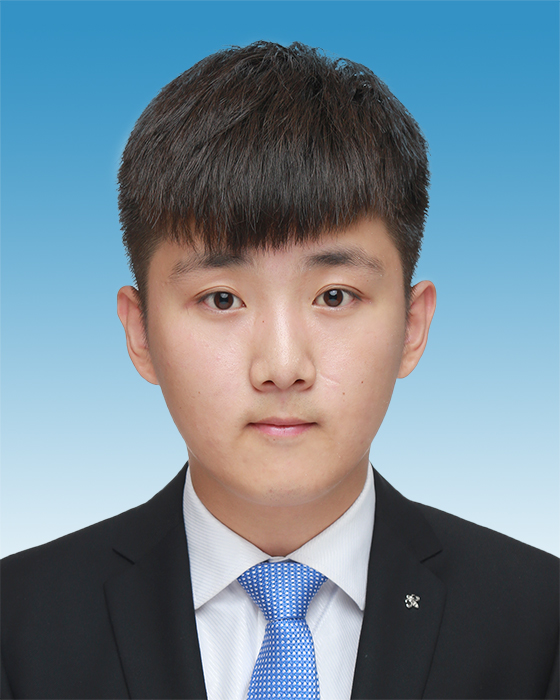}}]{Chunyu Qiang} received the M.S. degree from the University of Chinese Academy of Sciences, Beijing, China, in 2021, and the B.E. degree from the Liaoning University, Shenyang, China, in 2018. He currently serves as a researcher at the Audio Center of Kling AI Technology Department, Kuaishou Technology Co., Ltd., Beijing, China, while pursuing the Ph.D. degree at Tianjin University, Tianjin, China. His current research interests include speech synthesis, audio generation, and representation learning.
\end{IEEEbiography}

\begin{IEEEbiography}[{\includegraphics[width=1in,height=1.25in,clip,keepaspectratio]{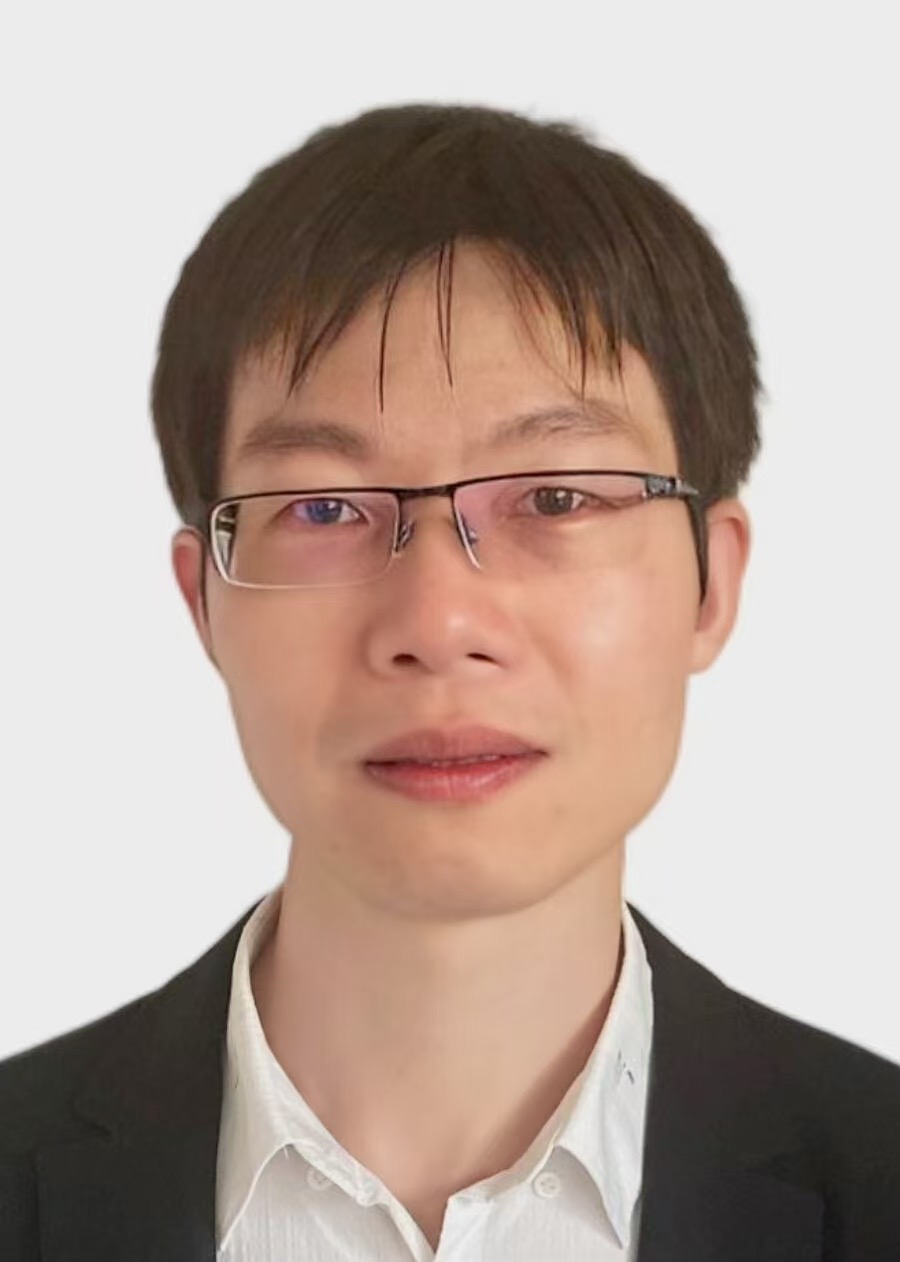}}]{Wang Geng} received his Ph.D. degree in pattern recognition and intelligent system from Institute of Automation, Chinese Academy of Sciences, Beijing, China, in 2017. He is currently a researcher at Huawei Co., Ltd., Beijing, China. His research interests include end to end speech understanding and generation.
\end{IEEEbiography}

\begin{IEEEbiography}[{\includegraphics[width=1in,height=1.25in,clip,keepaspectratio]{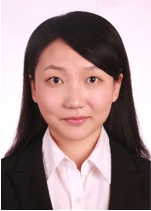}}] {Yi Zhao} received her M.S. degree from Tsinghua University, Beijing, China, in 2014, and her Ph.D. degree from the University of Tokyo, Tokyo, Japan, in 2018. She is currently a researcher at Kuaishou Technology Co., Ltd., Beijing, China. Her research interests include speech understanding and audio generation.
\end{IEEEbiography}

\begin{IEEEbiography}[{\includegraphics[width=1in,height=1.25in,clip,keepaspectratio]{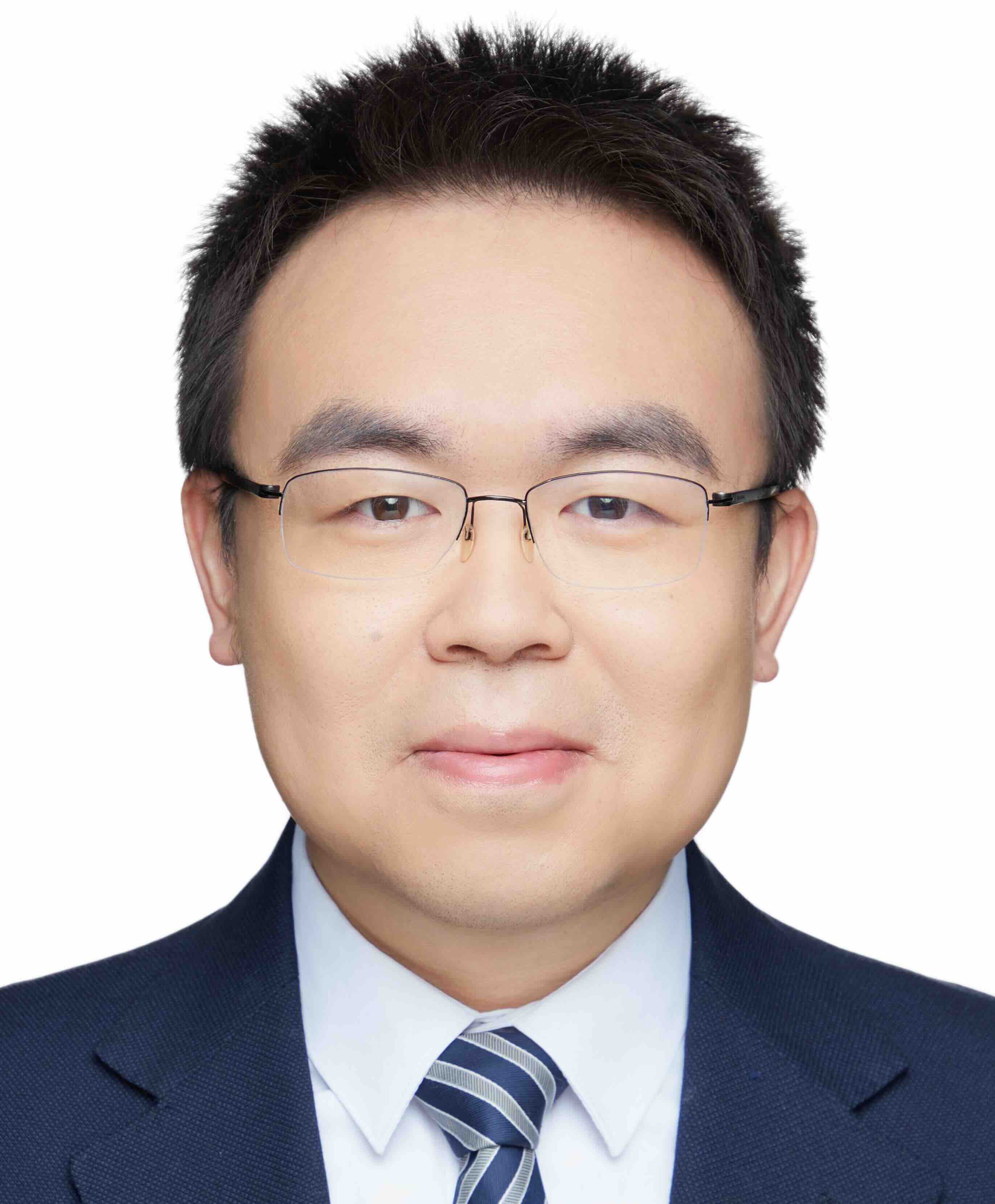}}]{Ruibo Fu} is an Associate professor in the National Laboratory of Pattern Recognition, Institute of Automation, Chinese Academy of Sciences, Beijing. He obtained B.E. from Beijing University of Aeronautics and Astronautics in 2015 and Ph.D. from the Institute of Automation, Chinese Academy of Sciences in 2020. His research interest is speech synthesis and transfer learning. He has published more than 20 papers in international conferences and journals such as ICASSP and INTERSPEECH and has won the best paper award twice in NCMMSC 2017 and 2019. He won the first prize in the personalized speech synthesis competition held by the Ministry of Industry and Information Technology twice in 2019 and 2020. He also won the first prize in the ICASSP2021 Multi-Speaker Multi-Style Voice Cloning Challenge (M2VoC) Challenge.
\end{IEEEbiography}

\begin{IEEEbiography}[{\includegraphics[width=1.1in,height=1.25in,clip,keepaspectratio]{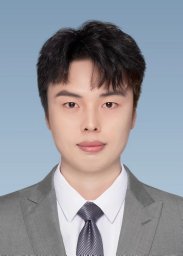}}]{Tao Wang} obtained his B.E. degree from the School of Control Science and Engineering, Shandong University (SDU), Jinan, China, in 2018. Subsequently, he pursued his Ph.D. at the National Laboratory of Pattern Recognition, Institute of Automation (NLPR), Chinese Academy of Sciences (CASIA), Beijing, China. During this period, his research focused on speech synthesis, voice conversion, machine learning, and transfer learning. Currently, he applies his research achievements to various fields, contributing to the practical implementation of relevant technologies and industry development.
\end{IEEEbiography}

\begin{IEEEbiography}[{\includegraphics[width=1in,height=1.25in,clip,keepaspectratio]{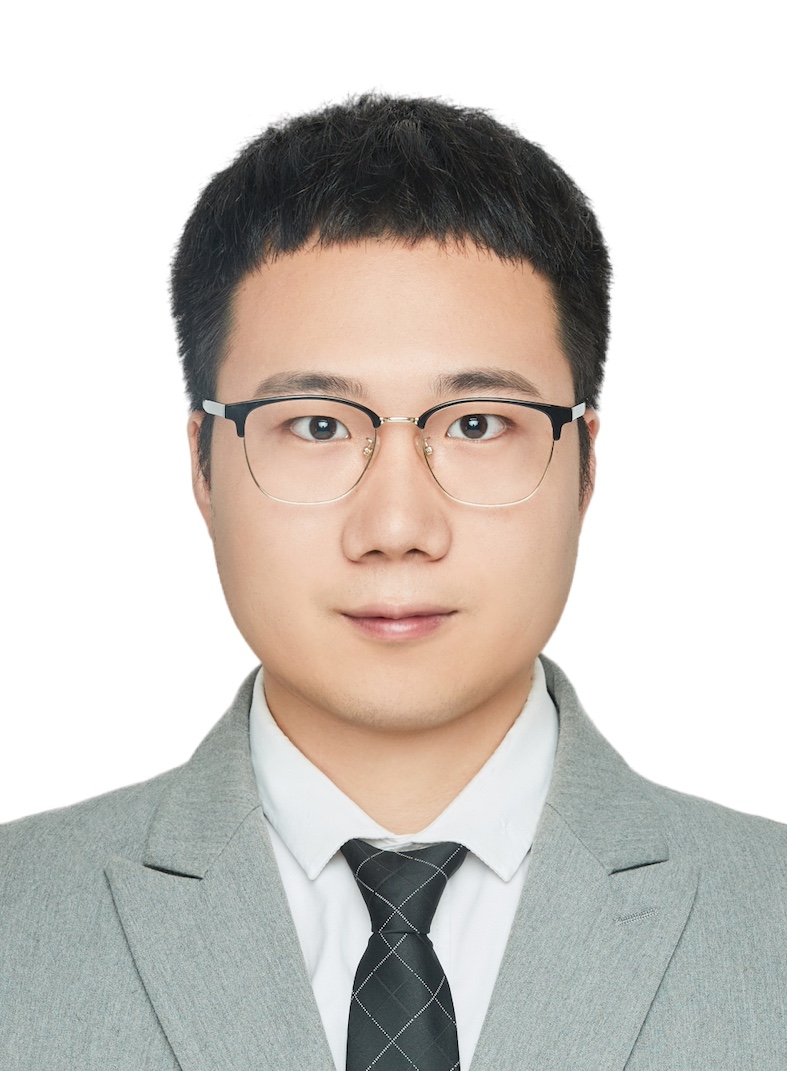}}]{Cheng Gong} received his B.S. and  M.S. degree from the Hohai University, Nanjing, China, in 2016 and 2019. He is currently working toward the Ph.D. degree in Tianjin Key Laboratory of Cognitive Computing and Application, Tianjin University, Tianjin, China. He is also an exchange Ph.D. candidate at the National Institute of Informatics, Japan, funded by China Scholarship Council (CSC). His research interests include expressive speech synthesis and multilingual speech synthesis. 
\end{IEEEbiography}

\begin{IEEEbiography}[{\includegraphics[width=1in,height=1.25in,clip,keepaspectratio]{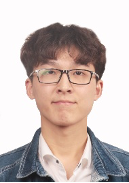}}]{Tianrui Wang} received his B.E. degree in Internet of Things Engineering from North University of China, Shanxi, China, in 2020. He obtained his M.S. degree in Information and Communication Engineering from Beijing Jiaotong University, Beijing, China, in 2023. He completed internships at China Mobile Research Institute in 2022 and at Microsoft Research Asia in 2023. Currently, he is pursuing a Ph.D. degree in Digital Information at Tianjin University, Tianjin, China. His current research interests include text-to-speech and self-supervised learning in speech processing. 
\end{IEEEbiography}

\begin{IEEEbiography}[{\includegraphics[width=1in,height=1.25in,clip,keepaspectratio]{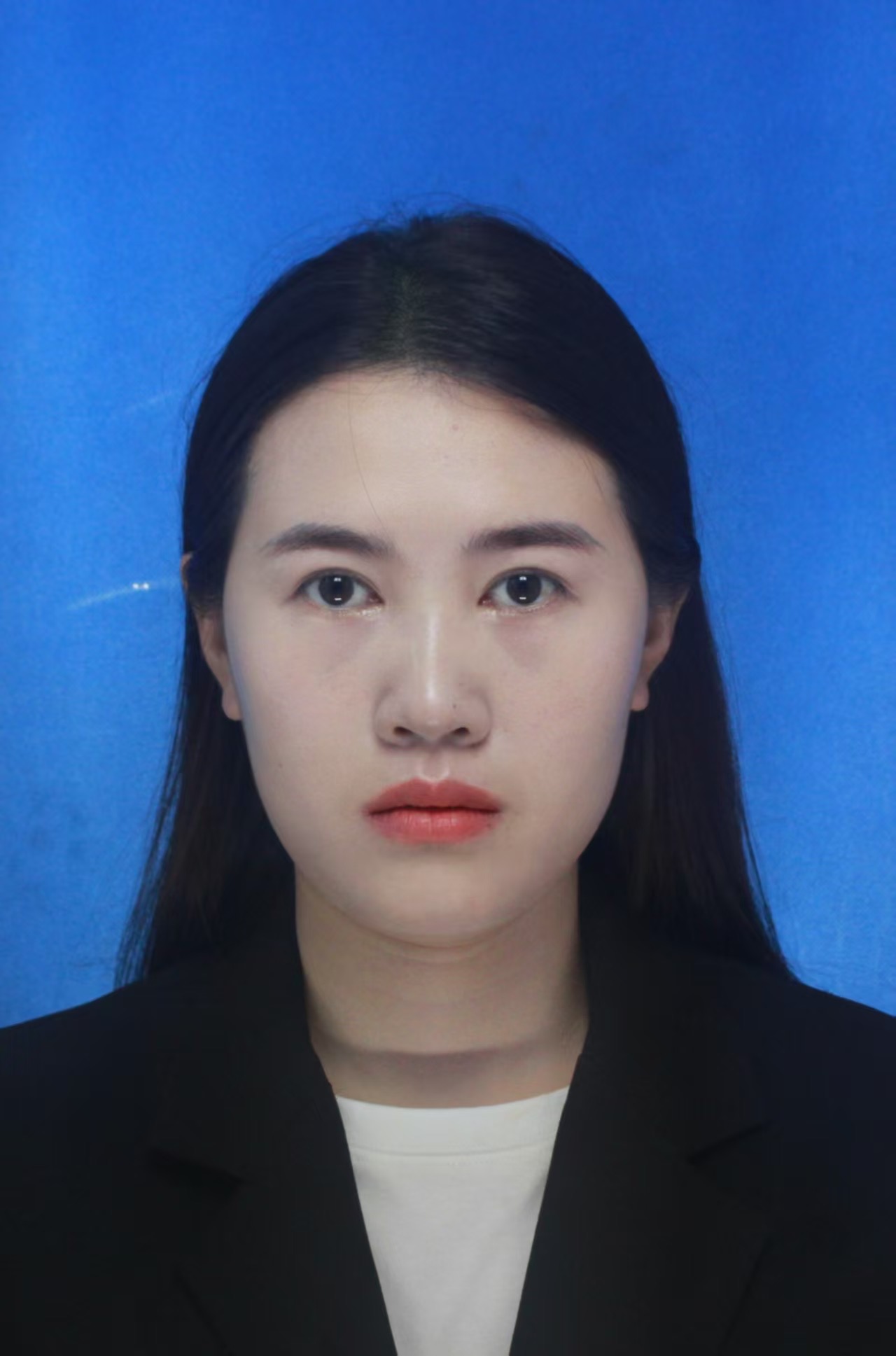}}]{Liu Qiuyu} received her B.Eng. degree from Henan University, Kaifeng, China, in 2022. From 2022 to 2025, she is a graduate student in the Department of Intelligence and Computing at Tianjin University in China, and join the Tianjin Key Laboratory of Cognitive Computing and Application. Her research interest is speech synthesis. 
\end{IEEEbiography}

\begin{IEEEbiography}
	[{\includegraphics[width=1in,height=1.25in,clip,keepaspectratio]{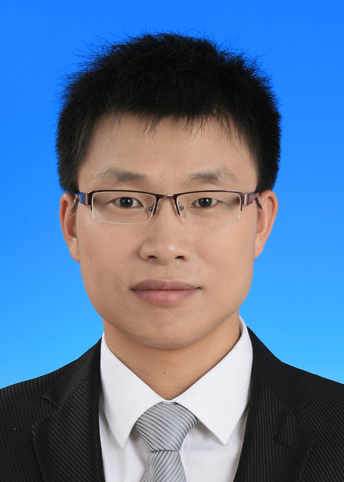}}]{Zhengqi Wen}
	(Member, IEEE) received the B.S. degree from the University of Science and Technology of China, Hefei, China, in 2008, and the Ph.D. degree from the Chinese Academy of Sciences, Beijing, China, in 2013, both in pattern recognition and intelligent system. He is currently an Associate Professor with the Beijing National Research Center for Information Science and Technology, Tsinghua University, Beijing, China. His current research interests include speech processing, speech recognition, and speech synthesis.
\end{IEEEbiography}

\begin{IEEEbiography}
	[{\includegraphics[width=1in,height=1.25in,clip,keepaspectratio]{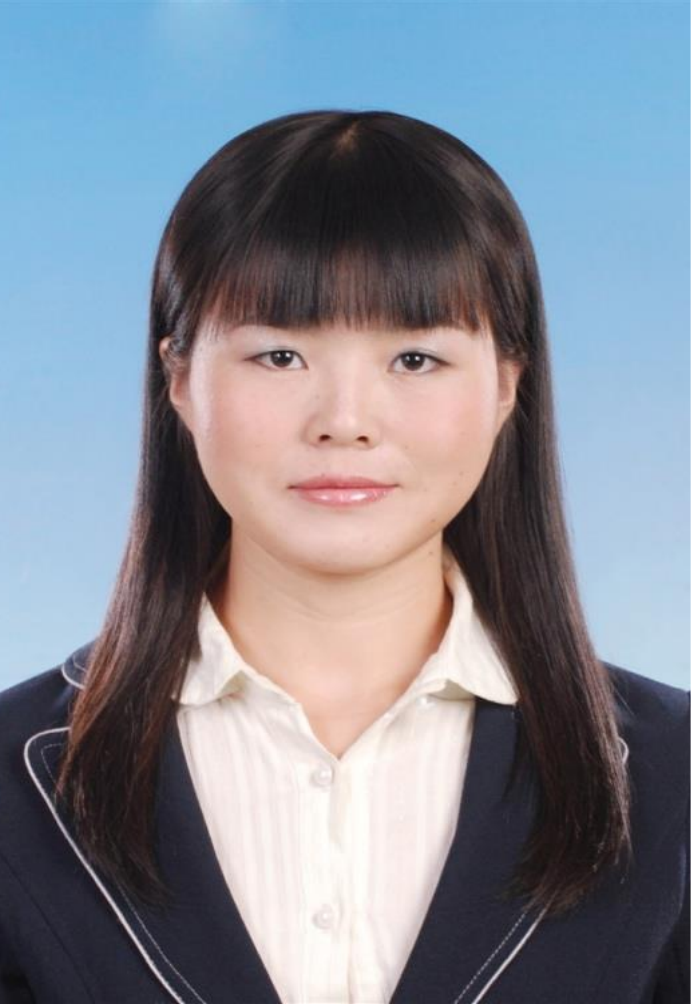}}]{Jiangyan Yi }
	(Member, IEEE) received the Ph.D. degree from the University of Chinese Academy of Sciences, Beijing, China, in 2018, and the M.A. degree from the Graduate School of Chinese Academy of Social Sciences, Beijing, in 2010. During 2011 to 2014, she was a Senior R\&D Engineer with Alibaba Group. During 2018 to 2024, she was an Associate Professor with the State Key Laboratory of Multimodal Artificial Intelligence Systems, Institute of Automation, Chinese Academy of Sciences. She is currently an Associate Researcher with the Department of Automation, Tsinghua University. Her research interests include speech signal processing, speech recognition and synthesis, fake audio detection, audio forensics, and transfer learning.
\end{IEEEbiography}

\begin{IEEEbiography}
	[{\includegraphics[width=1in,height=1.25in,clip,keepaspectratio]{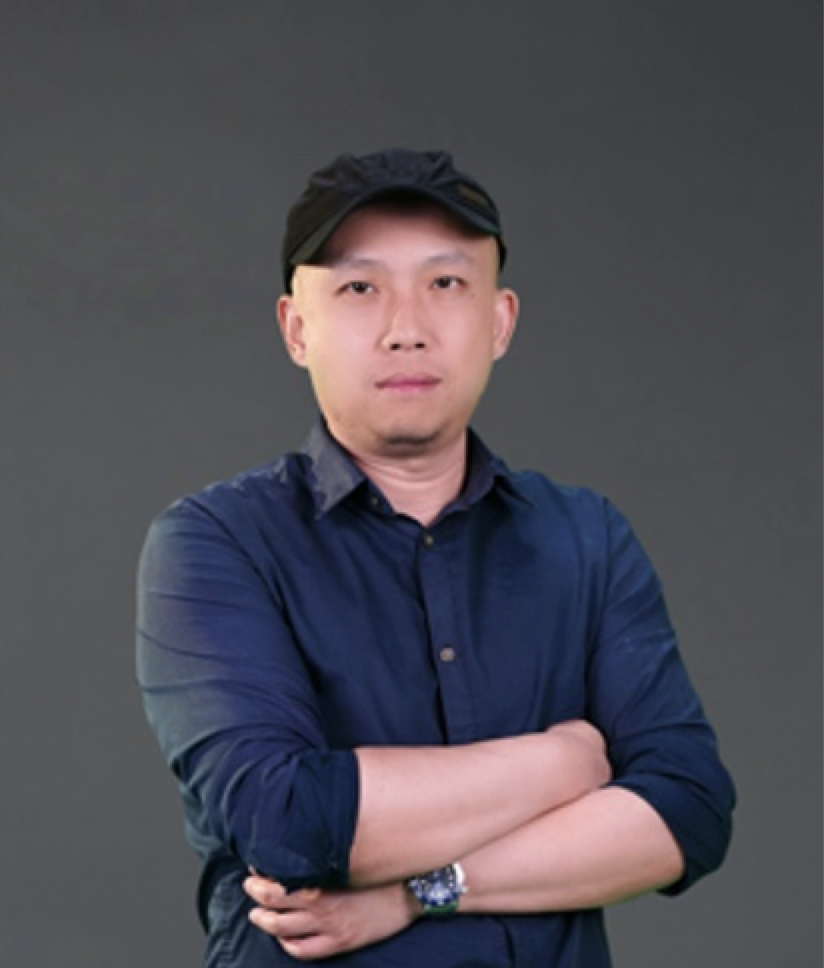}}]{Chen Zhang} received the B.E. degree from Tsinghua University, Beijing, China, in 2003, and the M.S. degree from the Institute of Acoustics, Chinese Academy of Sciences, Beijing, China, in 2006. He currently serves as a Director of the Audio Center at Kling AI Technology Department, Kuaishou Technology Co., Ltd., Beijing, China. His current research interests include speech synthesis, audio generation, speech recognition, and multimodal audio processing.
\end{IEEEbiography}

\begin{IEEEbiography}
	[{\includegraphics[width=1in,height=1.25in,clip,keepaspectratio]{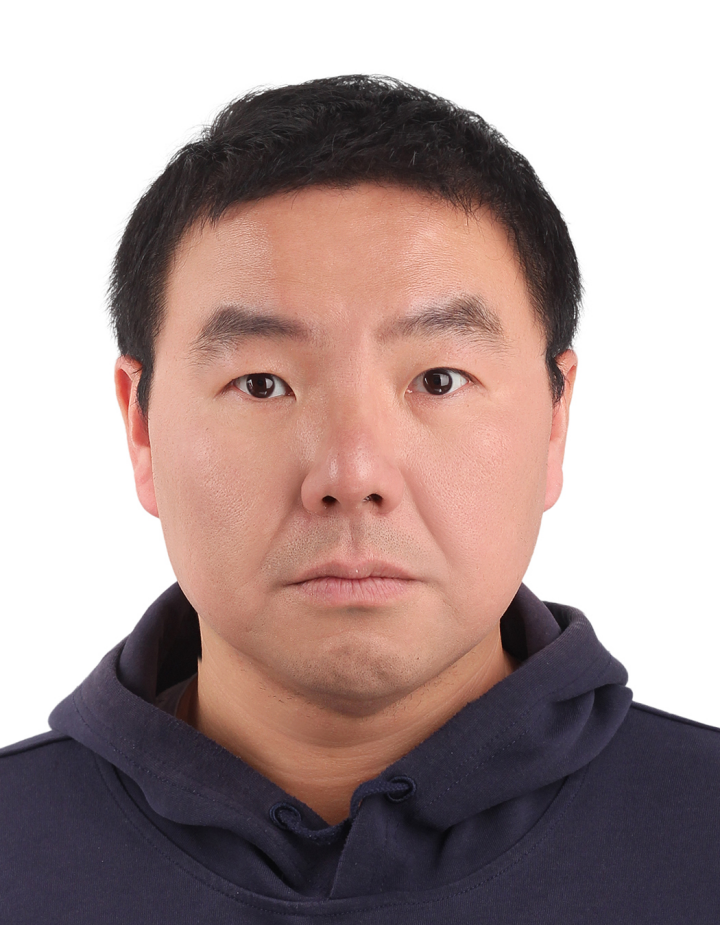}}]{Hao Che} received the B.S.degree in automatic control system from the Huazhong University of Science and Technology, Wuhan, China, in 2004, and the Ph.D. degree in pattern recognition and intelligent system fronthe Chinese Academy of Sciences, Beijing, China in 2015. He is currently working as an algorithm expert at China Mobile Migu Culture Technology Co., Ltd., Beijing. His research interests include speech processing speech recognition, human-computer interaction system.
\end{IEEEbiography}

\begin{IEEEbiography}[{\includegraphics[width=1in,height=1.25in,clip,keepaspectratio]{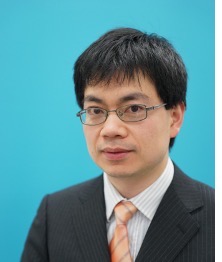}}]{Longbiao Wang} (Member, IEEE)
 received his Dr. Eng. degree from Toyohashi University of Technology, Japan, in 2008. He was an assistant professor in the faculty of Engineering at Shizuoka University, Japan from April 2008 to September 2012. He was an associate professor at Nagaoka University of Technology, Japan from Oct. 2012 to Aug. 2016. He is currently a professor, director of Tianjin Key Laboratory of Cognitive Computing and Application and vice dean of School of Artificial Intelligence at Tianjin University, China. His research interests include robust speech recognition, speaker recognition, acoustic signal processing and natural language processing. 
\end{IEEEbiography}

\begin{IEEEbiography}[{\includegraphics[width=1in,height=1.25in,clip,keepaspectratio]{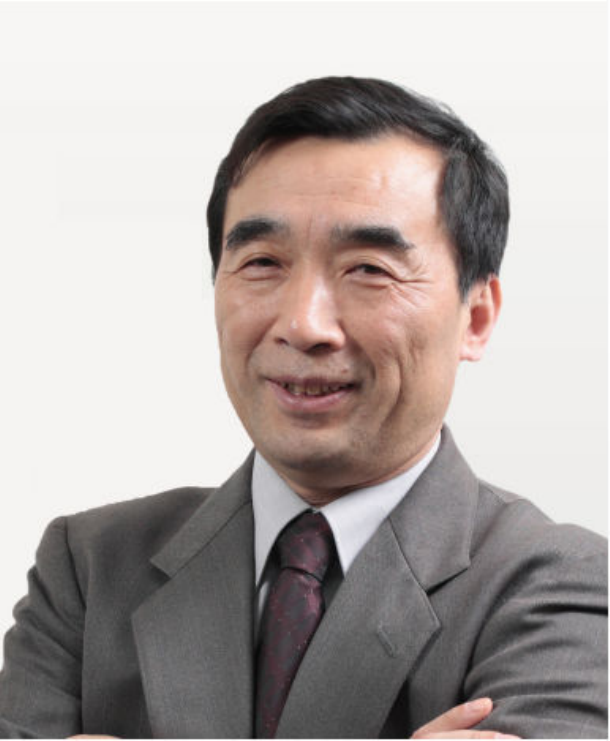}}]{Jianwu Dang}
 (Member, IEEE) graduated from Tsinghua Univ., China, in 1982, and got his M.S. degree at the same university in 1984. He worked for Tianjin Univ. as a lecture from 1984 to 1988. He was awarded the PhD degree from Shizuoka Univ., Japan in 1992. He worked for ATR Human Information
 Processing Labs., Japan, as a senior researcher from
 1992 to 2001. He joined the University of Waterloo,
 Canada, as a visiting scholar for one year from 1998.
 Since 2001, he has worked for Japan Advanced
 Institute of Science and Technology (JAIST) as a
 professor. He joined the Institute of Communication Parlee (ICP), Center of
 National Research Scientific, France, as a research scientist the first class from
 2002 to 2003. Since 2009, he has joined Tianjin University, Tianjin, China.
 His research interests are in all the fields of speech science including brain
 science, and speech signal processing. He built MRI-based bio-physiological
 models for speech and swallowing, and endeavors to apply these models on
 clinics.
\end{IEEEbiography}

\begin{IEEEbiography}
	[{\includegraphics[width=1in,height=1.25in,clip,keepaspectratio]{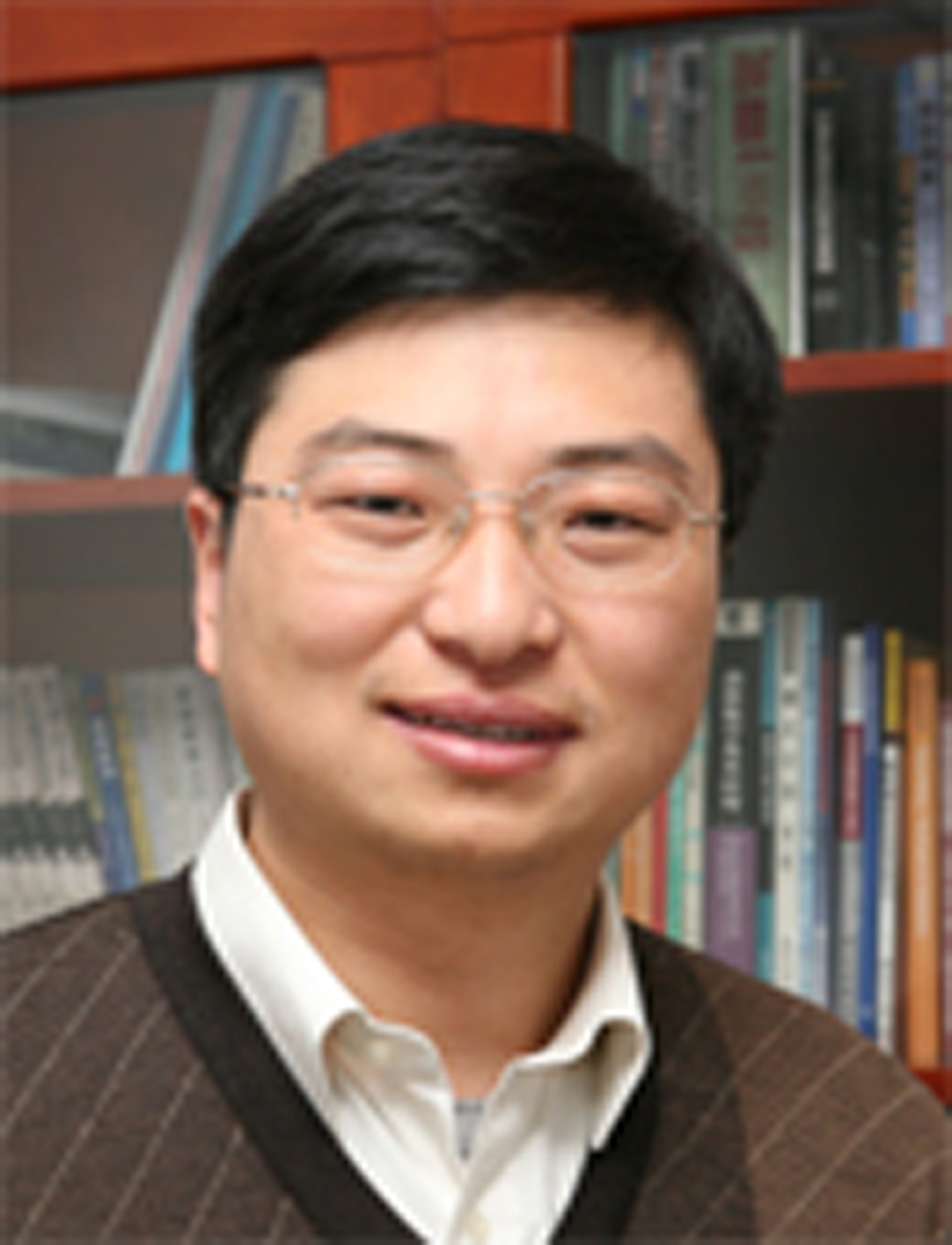}}]{Jianhua Tao}
	(Senior Member, IEEE) received the M.S. degree from Nanjing University, Nanjing, China, in 1996, and the Ph.D. degree from Tsinghua University, Beijing, China, in 2001. He is currently a Professor with Department of Automation, Tsinghua University, Beijing, China. He has authored or coauthored more than 300 papers on major journals and proceedings including the IEEE TASLP, IEEE TAFFC, IEEE TIP, IEEE TSMCB, Information Fusion, etc. His current research interests include speech recognition and synthesis, affective computing, and pattern recognition. He is the Board Member of ISCA, the chairperson of ISCA SIG-CSLP, the Chair or Program Committee Member for several major conferences, including Interspeech, ICPR, ACII, ICMI, ISCSLP, etc. He was the subject editor for the Speech Communication, and is an Associate Editor for Journal on Multimodal User Interface and International Journal on Synthetic Emotions. He was the recipient of several awards from important conferences, including Interspeech, NCMMSC, etc.
\end{IEEEbiography}

\vspace{11pt}

\vfill

\end{document}